\documentclass[aps,pra,showpacs,amsfonts,notitlepage,amssymb,amsmath,nofootinbib,superscriptaddress,twocolumn]{revtex4-1}
\usepackage[utf8]{inputenc}
\usepackage[english]{babel}
\usepackage[titletoc,toc,title]{appendix}
\usepackage{amsmath}
\usepackage{amsfonts}
\usepackage{amssymb}
\usepackage[colorlinks=true,linkcolor=blue,citecolor=blue,urlcolor=blue]{hyperref}
\usepackage{graphicx}
\usepackage{enumerate}
\usepackage{csquotes}
\usepackage[caption=false]{subfig}
\usepackage{dsfont}
\usepackage{xcolor}
\usepackage{bbold}
\usepackage{mathtools}
\usepackage{tensor}
\usepackage{gensymb}
\usepackage{ulem}

\begin{document}

\title{The DeLLight experiment to observe an optically-induced change of the vacuum index}
\date{\today}

\author{Scott~Robertson}
\affiliation{Universit\'e Paris-Saclay, CNRS/IN2P3, IJCLab, 91405 Orsay, France}
\author{Aur\'elie~Mailliet}
\affiliation{Universit\'e Paris-Saclay, CNRS/IN2P3, IJCLab, 91405 Orsay, France}
\author{Xavier~Sarazin}
\affiliation{Universit\'e Paris-Saclay, CNRS/IN2P3, IJCLab, 91405 Orsay, France}
\author{Fran\c{c}ois~Couchot}
\affiliation{Universit\'e Paris-Saclay, CNRS/IN2P3, IJCLab, 91405 Orsay, France}
\author{Elsa~Baynard}
\affiliation{Universit\'e Paris-Saclay, CNRS/IN2P3, IJCLab, 91405 Orsay, France}
\author{Julien~Demailly}
\affiliation{Universit\'e Paris-Saclay, CNRS/IN2P3, IJCLab, 91405 Orsay, France}
\author{Moana~Pittman}
\affiliation{Universit\'e Paris-Saclay, CNRS/IN2P3, IJCLab, 91405 Orsay, France}
\author{Arache~Djannati-Ata\"{i}}
\affiliation{Universit\'e Paris Diderot, CNRS/IN2P3, APC, Paris, France}
\author{Sophie~Kazamias}
\affiliation{Universit\'e Paris-Saclay, CNRS/IN2P3, IJCLab, 91405 Orsay, France}
\author{Marcel~Urban}
\affiliation{Universit\'e Paris-Saclay, CNRS/IN2P3, IJCLab, 91405 Orsay, France}

\begin{abstract}
Quantum electrodynamics predicts that the vacuum must behave as a nonlinear optical medium: the speed of light should be modified when the vacuum is stressed by intense electromagnetic fields.  
This optical phenomenon has not yet been observed.
The DeLLight (Deflection of Light by Light) experiment aims to observe the optically-induced index change of vacuum,
a nonlinear effect which has never been explored.
The experiment is installed in the LASERIX facility at IJCLab,
which delivers ultra-short intense laser pulses (2.5~J per pulse, each of 30~fs duration, with a 10~Hz repetition rate).
The proposal is to measure the refraction of a probe laser pulse when crossing a transverse vacuum index gradient, produced by a very intense pump pulse. 
The refraction induces a transverse shift in the intensity profile of the probe, whose signal is amplified by a Sagnac interferometer. 
In this article, we describe the experimental method and setup, and present the complete theoretical calculations for the expected signal. 
With a minimum waist at focus of $5 \,\mu$m (corresponding to a maximum intensity of $\sim 3 \times 10^{20}$~W/cm$^2$), and with the nonlinear vacuum index derived from QED, the expected refraction angle is 0.13~prad. 
First results of the interferometer prototype are presented.
It is shown that an extinction factor $\mathcal{F} = 0.4 \times 10^{-5}$ (corresponding to a signal amplification factor of 250) and a spatial resolution $\sigma_y = 10$~nm are achievable. 
The expected signal is then about 15~pm, and could be observed at a 5-sigma confidence level with about one month of collected data.
\end{abstract}

\maketitle

\section{Introduction
\label{sec:introduction}}

The classical electromagnetic vacuum is described as a linear optical medium. The speed of light in vacuum, $c$, and the related vacuum 
permeability $\mu_0$ and 
permittivity $\epsilon_0$ are universal constants. The Maxwell equations for the electromagnetic fields are linear and the vacuum constants $c$, $\mu_0$ and $\epsilon_0$ do not depend on externally applied fields. 
In media, however,
the dependence of the optical index on the electromagnetic fields has been known since the 19th century, with Faraday's discovery of circular
birefringence induced by an external magnetic field in the direction of propagation (Faraday effect)~\cite{Faraday-1846}, 
and Kerr's discovery of linear birefringence induced by a transverse electric field (Kerr effect)~\cite{Kerr-1875}. 
Both effects can be interpreted as polarization-dependent changes of the refractive index, and while the former is linear in the field strength $B$, the latter is proportional to the intensity $I \propto E^{2}$.  
The optical Kerr effect, in which the electric field responsible for the index change is due to light itself, has been extensively studied and measured 
thanks to the availability of high-intensity lasers in the last twenty years~\cite{Bloembergen-1996}.

By analogy with the situation in media, we may ask whether the vacuum is also a nonlinear optical medium, i.e., whether the vacuum 
index is enhanced when the vacuum is stressed by intense externally applied fields. 
Born and Infeld~\cite{Born-Infeld} were the first to introduce nonlinear electrodynamics terms in vacuum by assuming an absolute maximum of the electric field in order to regularize the electromagnetic field of a point charge, and thus to obtain a finite electromagnetic mass of the electron equal to its observed mass. 
Soon after, Euler, Kockel and Heisenberg~\cite{Euler-Heisenberg} derived an effective nonlinear electromagnetic field theory with nonlinear terms induced by the coupling of the fields with the electron-positron virtual pairs in vacuum. 
This is described by the so-called Euler-Heisenberg nonlinear Lagrangian, and was later reformulated as photon-photon scattering (four-waves interactions) in Quantum Electrodynamics (QED). 

The first observation of photon-photon scattering was 
obtained at the Stanford Linear Accelerator Center (SLAC) by measuring the collision and fusion between a high energy (GeV) gamma and several laser photons~\cite{SLAC} to produce an electron-positron pair (multiphoton Breit-Wheeler reaction). 
More recently, 
high energy gamma-gamma pair emission from virtual gamma-gamma scattering in ultra peripheral Pb-Pb collisions has been observed by the  ATLAS and CMS detectors at the Large Hadron Collider (LHC)~\cite{ATLAS,CMS}.  
Both cases involve inelastic high-energy photon-photon scattering, described by a four-photons Feynman diagram. The vacuum appears in the exchange of an electron-positron virtual pair, which can even become real in the case of the Breit-Wheeler process. 
However, in these scattering processes, there is no modification of the optical properties of the vacuum, i.e. no modification of the vacuum electromagnetic constants $c$, $\epsilon_0$ and $\mu_0$.

Another approach is to search for a direct manifestation of a nonlinear optical effect in vacuum, 
a coherent phenomenon corresponding to a pure undulatory process at large scale and treated classically in the long-wavelength limit.
This effect differs drastically from the inelastic photon-photon scattering since it corresponds to a nonlinear modification of the fundamental level of the electromagnetic vacuum, with a change of the vacuum speed of light at macroscopic scale. 
Experimental efforts have mainly involved testing vacuum magnetic birefringence in the presence of an external magnetic field of 2-30 Tesla~\cite{PVLAS,BMV,OVAL,QandA}. 
This process is predicted by Euler-Heisenberg and often referred to 
as the Cotton-Mouton effect (in vacuum). 
As yet, no signal has been observed, and the experimental uncertainty is about one order of magnitude above the predicted QED value~\cite{PVLAS}.
Several theoretical works have considered the possibility of using high-intensity laser pulses to increase the strength of the field~\cite{Aleksandrov-1985,Heinzl-2006,King-Heinzl-2016,Schlenvoigt-2016,Shen-2018}.
However, it is important to note that this approach is insensitive to Born-Infeld type models, since these predict no vacuum birefringence. 

Alternatively, one can adopt a complementary approach that is independent of the occurrence of birefringence, by directly exploiting  
the change in the refractive index induced by the nonlinearity. One then expects 
refraction of a light beam in response to this index change.  
A test of refraction due to nonlinear electrodynamical effects was last performed in 1960 by Jones~\cite{Jones1,Jones2}, who looked for the deviation of a light beam passing through a transverse static magnetic field of about 1~T. Results indicated that the deflection of the light beam was less than 0.5~prad, while the deflection angle predicted by the Euler-Heisenberg model was almost 10 orders of magnitude below this limit. 
As for birefringence-based experiments, there are several theoretical works considering the use of high-intensity laser pulses to achieve larger index changes; such proposals include the emission of Cherenkov radiation by a charged particle~\cite{Dremin-2002,Macleod-2019,Bulanov-2019} and nonlinear wave propagation~\cite{Kadlecova-2019,Bulanov-2020}.

The DeLLight (Deflection of Light by Light) project proposes a similar test to that of Jones, 
using the
much stronger electromagnetic field contained in high-intensity and ultra-short focused 
laser pulses. 
It proposes to test the nonlinear nature of vacuum by measuring the refraction of a probe pulse crossing an index gradient engendered by an energetic (Joule) and ultra-short (femtosecond) laser pulse. 
This is formally similar to the induction of an index change by the optical Kerr effect in a medium. 
The DeLLight idea was first 
proposed in~\cite{DeLLight-2016} using a simplified theoretical model to calculate the expected signal and a simplified experimental model to roughly estimate the sensitivity. 

In this article we propose a new experimental setup, described in Section~\ref{sec:setup}, and we present in Section~\ref{sec:expected-signal-sensitivity} the complete theoretical and numerical calculations for the expected signal. We then present in Section~\ref{sec:prototypes} preliminary results of the experimental performances achieved with the DeLLight prototypes.

\section{Description of the proposed DeLLight experimental method
\label{sec:setup}}

In this section we describe the main concepts involved in the DeLLight project.  We then give a detailed description of the experimental setup.

The DeLLight experiment is installed in the  LASERIX facility (at IJCLab, Orsay), which delivers ultra-high intense laser pulses with a repetition rate of 10~Hz. The current energy is 1.5~J per pulse with a duration of $T_0 = 40$~fs (FWHM in intensity). The laser will be upgraded in the coming years to reach 2.5~J per pulse with a duration of $T_0 = 30$~fs, corresponding to a maximum
intensity of $3 \times 10^{20}$~W/cm$^2$ 
(or a maximum $B \sim 10^{5} \, {\rm T}$, with $E = c B$) for
a minimum waist at focus of $5 \mu$m. 
The DeLLight principle, illustrated by the schema in Figure~\ref{fig:pump-probe-refraction}, is to cross two counter-propagating ultra-short laser pulses at their focus point. 
One pulse is very intense and is called the {\it pump}; the other is of much lower intensity and is called the {\it probe}.  
The pump engenders a propagating refractive index profile 
which, by analogy with the optical Kerr effect in a medium, can be written as
$\delta n = n_{2} I_{\rm pump}$, where $I_{\rm pump}$ is its intensity 
(averaged over the rapid oscillations of the carrier wave) and where, in a Lorentz-invariant model with local photon-photon interactions, 
the nonlinear index $n_{2}$ is given by~\cite{Robertson-2019}
\begin{equation}
n_{2} = n_{2,{\rm max}} \, r_{\rm pol} \, {\rm cos}^{4}\left(\frac{\theta_{\rm tilt}}{2}\right) \,.
\label{eq:n2}
\end{equation}
In the Euler-Heisenberg model derived from QED, the optimum value of the nonlinear index is 
\begin{eqnarray}
n_{2,{\rm max}} = n_{2,{\rm QED}} &=& \frac{56}{45} \alpha^{2} \frac{\hbar^{3}}{m_{e}^{4} c^{6}}  \nonumber \\ 
&=& 1.55 \times 10^{-33} \, {\rm cm}^{2}/{\rm W}, 
\end{eqnarray}
where $\alpha \approx 1/137$ is the fine structure constant and $m_{e}$ is the mass of the electron.
The factor $r_{\rm pol} \in \left[ \frac{4}{7}, 1 \right]$ accounts for the birefringence of the 
model and its dependence 
on the polarization state of the pump.  We shall here assume the optimized case $r_{\rm pol} = 1$, which occurs when the pump and probe are each linearly polarized, with their electric (or magnetic) fields orthogonal to each other.  Finally, $\theta_{\rm tilt}$ is the tilt angle between the propagation directions of pump and probe, defined to be zero when they are exactly counter-propagating (i.e., when the collision is ``head-on'').
The expected Euler-Heisenberg value of the vacuum nonlinear index 
$n_{2,{\rm QED}}$ is orders of magnitude smaller than the Kerr index values measured in silica, $\simeq 2 \times 10^{-16} \, {\rm cm}^{2}/{\rm W}$~\cite{Kerr-index-SiO2} and in air $\simeq 3 \times 10^{-19} \, {\rm cm}^{2}/{\rm W}$~\cite{Kerr-index-air}.

Consider now the weak probe wave crossing the pump pulse.  It will ``see'' and react to the refractive index profile just described.  In particular,  in the interaction area where the pump intensity is rather highly concentrated (as can be achieved by strong focusing), the index profile will have a significant gradient in the transverse directions, and this will tend to bend the rays of the probe towards regions of higher index/intensity.
The wavefronts of the probe pulse are rotated (refracted), as illustrated in Figure~\ref{fig:pump-probe-refraction}. 

As will be shown in Sec.~\ref{sec:expected-signal-sensitivity},
for pump parameters achievable with the LASERIX facility the expected deflection angle is $\approx 0.1$~picoradian. 
This deflection is challengingly 
small, so the experimental setup is designed such as to amplify the signal, bringing the deflection into an observable range.  This is the Sagnac interferometer, in which an initial probe wave is incident on a 50/50 beamsplitter, generating two daughter pulses, named Sagnac pulses, which propagate along a common path in opposite directions and are then recombined at the same beamsplitter.  
One of the outputs of the 
beamsplitter is a ``dark port'' where the two Sagnac pulses have nearly equal amplitude and a relative phase close to $\pi$, so that they interfere destructively.  
The degree of alignment is characterized by the {\it extinction factor} $\mathcal{F}$, defined as $\mathcal{F}=I_{\mathrm{out}}/I_{\rm in}$, where $I_{\rm in}$ is the incident intensity and $I_{\mathrm{out}}$ is the intensity in the dark output of the interferometer.  
A perfectly aligned Sagnac interferometer thus has $\mathcal{F} = 0$.  
What makes this useful is that 
any perturbation of one Sagnac pulse with respect to the other during their propagation around the common path of the interferometer will in turn perturb the destructive interference at the dark port, and the smaller is the value of $\mathcal{F}$, the larger (relatively speaking) will be the change observed in the output at the dark port.  This is the principle behind the amplification by the Sagnac interferometer.
Measurements of the deflection of a laser beam through amplification with a Sagnac interferometer have already been developed with continuous laser beams in the search for gravitational anomalies at short distance, and angular deflections of a mirror down to the sub-picoradian scale have been measured with this technique~\cite{rochester,seattle}. 
However, no refraction measurement has yet been performed in femtosecond pulsed mode with a Sagnac interferometer.

The proposed experimental setup of the DeLLight project is shown in Figure~\ref{fig:setup}.  
The pump pulse, with a polarisation $s$, is focused by an off-axis parabolic mirror (OAP) in the interaction area.  A much weaker pulse (few tens of $\mu$J), with a perpendicular polarisation $p$, is sent into a Sagnac interferometer via a 50/50 beamsplitter (BS-2), generating two daughter pulses (Probe and Ref) that circulate in opposite directions around the interferometer. The interferometer is in a right-angled isosceles triangle configuration, formed by the beamsplitter and two mirrors (M1 and M2). 
The length of the longest interferometer arm between M1 and M2 is about 1 meter.
Both counter-propagating pulses are focused in the interaction area via two optical lenses (L1 and L2) of focal length $f$ inserted in the interferometer between the two mirrors. 
One pulse (Probe) refers to the probe pulse that is counter-propagating with respect to the pump, and a delay stage ensures the time coincidence of the arrival of this probe pulse with that of the pump pulse in the interaction area.  
The second pulse (Ref) is not in time coincidence and does not overlap with the pump. This pulse is therefore unaffected by the pump and refers to the reference pulse. 
The focus axis of the pump is transversally (vertically) shifted with respect to the focus axis of the probe, thus engendering an impact parameter $b$ so that the perturbation of the probe is asymmetric and the mean deflection is non-zero.
In the absence of the pump, the two counter-propagating probe and reference pulses are phase shifted by $\pi$ in the dark output of the interferometer, where a CCD camera measures the transverse position of the residual intensity profile.
When the pump pulse is present and interacts with the probe pulse, the wave fronts 
of the probe are refracted by the induced vacuum index gradient, while those of the reference pulse are unaffected.  
After recollimation, the refracted probe pulse is transversally (vertically) shifted with respect to the unrefracted reference pulse by an average distance $\langle \delta y \rangle_{\rm direct} = \langle \delta\theta \rangle \times f$, where $\langle \delta\theta \rangle$ is the average deflection angle of the refracted probe pulse. 
The interference of the probe and reference pulses in the output of the interferometer produces a transverse vertical displacement $\langle \delta y \rangle_{\rm Sagnac}$ of the barycenter $\bar{y}$ of the residual intensity profile, which is measured by the CCD camera.
As mentioned above, the advantage of the proposed interferometric method is an amplification of the measured signal $\langle \delta y \rangle_{\rm Sagnac}$, as compared to the would-be signal $\langle \delta y \rangle_{\rm direct}$ when using a standard direct pointing method. 
We will see that the amplification factor, defined as $\mathcal{A} = \langle \delta y \rangle_{\rm Sagnac}/\langle \delta y \rangle_{\rm direct}$, scales as $\mathcal{F}^{-1/2}$.

The presence of beam pointing fluctuations between successive laser pulses induces fluctuations of the position of the intensity profile in the output of the interferometer of typical magnitude much larger than the expected 
displacement signal $\langle \delta y \rangle_{\rm Sagnac}$. However, as illustrated in Figure~\ref{fig:schema-beampointingfluctuation}, an essential  advantage of the Sagnac interferometer is that the interference pattern and the extinction factor in the dark output are unmodified in the presence of beam pointing fluctuations, assuming the phase noise defined in Sec.~\ref{subsec:beampointing} is negligible.  
Fluctuations of the beam pointing produce a simple translation of the intensity profile on the CCD camera, which can be measured and suppressed by monitoring 
the position of the intensity profile with respect to the back-reflections from the rear side of the beamsplitter.
This is discussed in detail in Section~\ref{subsec:beampointing}.

\section{Expected signal and sensitivity}
\label{sec:expected-signal-sensitivity}

In this section we present the results of analytical calculations concerning the interaction between pump and probe, and the resulting deflection of the barycenter of the intensity profile.

\subsection{Deflection of probe rays by the effective $\delta n$}
\label{sub:deflection_rough}

As mentioned in Sec.~\ref{sec:setup}, 
when the weak probe pulse collides with the pump pulse, it engenders a vacuum index gradient which tends to bend the rays of the probe towards regions of higher index/intensity.
This is a direct consequence of Fermat's principle: 
taking $z$ as the propagation direction of the probe, and $x$ and $y$ as the transverse directions, the optical path length associated with a ray of the probe is
\begin{eqnarray}
\int \left(1+\delta n\right) {\rm d}s &=& \int \left(1 + \delta n\right) \sqrt{{\rm d}x^{2} + {\rm d}y^{2} + {\rm d}z^{2}} \nonumber \\
&\approx& \int \left( 1 + \delta n + \frac{\left(x^{\prime}\right)^{2}+\left(y^{\prime}\right)^{2}}{2} \right) {\rm d}z \,,
\end{eqnarray}
where the prime denotes differentiation with respect to $z$ along the ray trajectory, and where in the second line we have assumed both $\delta n \ll 1$ and $\left(x^{\prime}\right)^{2}+\left(y^{\prime}\right)^{2} \ll 1$ so that higher-order terms can be neglected.  This integral is minimized by the actual trajectory, and the problem is formally equivalent to a variational problem in classical mechanics, with $z$ playing the role of time, $\left(\left(x^{\prime}\right)^{2}+\left(y^{\prime}\right)^{2}\right)/2$ the kinetic energy, and $-\delta n$ the potential energy.  
Since we assume the deflections to be small, $x^{\prime}$ and $y^{\prime}$ are the angles $\theta_{x}$ and $\theta_{y}$ between the tangent vector of the trajectory and the $z$-axis.
Therefore, the corresponding Euler-Lagrange equations can be written as
\begin{equation}
\frac{{\rm d}\vec{\theta}}{{\rm d}z} = \nabla_{\perp} \left(\delta n\right) \,, \quad {\rm where} \quad \vec{\theta} = \left(\theta_{x}, \theta_{y}\right) = \left( \frac{{\rm d}x}{{\rm d}z} \,,\, \frac{{\rm d}y}{{\rm d}z} \right) \,.
\label{eq:deflection_rate}
\end{equation}

This allows us to determine the expected scaling of the accumulated deflection angle with the parameters characterizing the pump pulse.  We assume an axisymmetric pulse characterized by two scales: the transverse width $W_{0}$ and the duration $T_{0}$.  The index perturbation $\delta n = n_{2} \, I_{\rm pump}$ will scale as $n_{2}\,\mathcal{E}/T_{0}/\left( \pi W_{0}^{2} \right)$,
where $\mathcal{E}$ is the energy of the pump pulse.  The transverse gradient of $\delta n$,  equal to $\partial\left(\delta n\right)/\partial y$, will scale as $n_{2} \, I_{\rm pump}/W_0 \sim n_{2} \, \mathcal{E}/T_{0}/(\pi W_{0}^{3})$, and Eq.~(\ref{eq:deflection_rate}) tells us that this is equal to the deflection rate in the propagation direction $z$.  Integrating over $z$, and noting that the longitudinal size of the pump pulse is of order $c\,T_{0}$, we find 
\begin{equation}
\delta\theta \sim n_{2} \, c \, \frac{\mathcal{E}}{\pi W_{0}^{3}} \,.
\label{eq:delta-theta_scaling}
\end{equation}
This is a rough estimate; it does not account for the non-uniformity of $\partial\left(\delta n\right)/\partial y$ within the pump pulse, nor the spatial extent of the probe, which will itself be a beam rather than an individual ray.  But it does allow us to extract an expected order of magnitude for the deflection angle: letting $n_{2} = n_{2,{\rm max}} = 1.55 \times 10^{-33} \, {\rm cm}^{2}/{\rm W}$, 
$\mathcal{E} = 2.5\,{\rm J}$, and $W_{0} = 5\,\mu{\rm m}$, we find $\delta\theta \sim 0.3 \times 10^{-12} \, {\rm rad} = 0.3\,{\rm prad}$.  Since this approximates the probe as a ray of infinitesimal width, we can expect the average deflection of a spatially extended probe pulse to be smaller, so that this value must be an overestimate.
Indeed, we shall see below that taking the finite extension of the probe into account tends to reduce this estimate.

\subsection{Mean deflection using Gaussian pulses}
\label{sub:deflection}

The finite spatial extent of both the pump and the probe are taken into account to get a more accurate prediction.
For simplicity here, we assume the pulses to be exactly counter-propagating (i.e., $\theta_{\rm tilt} = 0$).
The general case, in which the tilt angle $\theta_{\rm tilt}$ is non-zero, is presented in Appendix~\ref{Appendix-A}.

For definiteness and simplicity, let us assume that the probe and pump intensity profiles are separable into transverse $F_{\perp}(x,y)$ and longitudinal $G_{z}(z)$ Gaussian profiles:
\begin{equation}
I(x,y,z) = F_{\perp}(x,y) \, G_{z}(z)
\label{eq:intensity_separable}
\end{equation}
with $ G_{z}(z) = e^{-\frac{4 \ln2 }{T_0^2}z^2} $ and
\begin{equation}
F_{\perp}(x,y) = \left\{ \begin{array}{ccll} i_{\perp}(x,y) & = & I_{\rm in} \, e^{- \frac{2}{w_{0}^{2}} \left(x^{2}+y^{2}\right)} & {\rm for \, probe} \\
I_{\perp}(x,y) & = & I_{0} \, e^{- \frac{2}{W_{0}^{2}} \left(x^{2}+\left(y - b\right)^{2}\right)} & {\rm for \, pump.} \end{array} \right.
\label{eq:intensity_transverse}
\end{equation}
Here $T_0$ is the duration of the pulse (FWHM), while the length scales $W_{0}$ and $w_{0}$ are the standardly defined waists at focus of the pump and probe, respectively. We have included an impact parameter, $b$, between the trajectories of the pump and probe pulses.  As mentioned in Sec.~\ref{sec:setup}, this is required because a non-zero deflection of the barycenter requires some asymmetry in the response of the probe to the presence of the pump, whereas it will be exactly symmetrical if $b = 0$.  We can also justify a non-zero impact parameter by noting (see Eq.~(\ref{eq:deflection_rate})) that the rate of deflection is equal to the {\it gradient} of $\delta n$ in the transverse directions, so we wish for the probe to be centered on a region where the pump intensity is strongly varying.

We can also consider the pulses to be collimated in the interaction region. 
Indeed, for a Gaussian beam propagation, the beam width varies longitudinally from the focus point as $w(z) = w_0 \sqrt{1+(z/z_R)^2}$, where $z$ is the longitudinal position along the beam measured from the focus, and $z_R=\pi w_0^2/\lambda$ is the Rayleigh length. 
For a pulse duration $T_0 = 30$~fs, the longitudinal size of the interaction region is of order $z_l = c\,T_{0} = 9 \,\mu{\rm m}$. 
For a waist at focus $w_{0} = 5\,\mu{\rm m}$ and a wavelength $\lambda = 800\,{\rm nm}$, the Rayleigh length is $z_R \simeq 100 \ \mu{\rm m}$. 
Thus, the relative variation of the beam width during the interaction 
is $(w(z= c\,T_{0})-w_0)/w_0 \simeq (z_l/z_R)^2/2 = 4\times10^{-3}$, so the approximation of collimated pulses in the interaction region is valid.

The fact that the total deflection is an integrated effect means that $g_{z}(z)$ drops out of the result, and to each transverse position $(x,y)$ we may assign a local deflection angle $\delta\theta_{y}(x,y)$.
Note that, since the focus axes of pump and probe are considered to be shifted along $y$, the probe ``sees'' a symmetric $\delta n$ in $x$, so the mean of $\delta\theta_{x}$ will vanish; it is for this reason we restrict our attention to $\delta\theta_{y}$.  
As per Eq.~(\ref{eq:deflection_rate}), the deflection is found by integrating $\partial_{y}\left(\delta n\right)$ over $z$: $\delta\theta_{y} = \partial_{y}\left(\int {\rm d}z \, \delta n\right)$.  
Writing $\delta n = n_{2,{\rm max}} \times I_{\rm pump} = n_{2,{\rm max}} \times I_{\perp}\,G_{z}$ and integrating over $z$, we find the local deflection angle $\delta\theta_{y}(x,y)$ to be:
\begin{eqnarray}
\delta\theta_{y}(x,y) &=& \partial_{y}\left[ \frac{1}{\pi} \, c \, n_{2,{\rm max}} \, \frac{\mathcal{E}}{W_{0}^{2}} \, e^{-\frac{2}{W_{0}^{2}}\left(x^{2}+\left(y - b\right)^{2}\right)} \right]  .
\label{eq:Gaussian_theta_intensity}
\end{eqnarray}
Here, $\mathcal{E}$ is the total energy of the pump pulse.

The mean deflection $\langle \delta\theta_{y} \rangle$ is found by averaging $\delta\theta_{y}(x,y)$ over $x$ and $y$, the transverse form of the probe intensity acting as a weight function:
\begin{equation}
\langle \delta\theta_{y} \rangle = \frac{\iint {\rm d}x\,{\rm d}y\, \delta\theta_{y}\left(x,y\right) \, i_{\perp}\left(x,y\right)}{\iint {\rm d}x\,{\rm d}y\, i_{\perp}\left(x,y\right)} \,.
\label{eq:mean_deflection}
\end{equation}
Substituting~(\ref{eq:Gaussian_theta_intensity}) into~(\ref{eq:mean_deflection}), we find:
\begin{equation}
\langle \delta\theta_{y} \rangle = \langle \delta\theta_{y} \rangle_{\rm max} \, \frac{b}{b_{\rm opt}} \, e^{ \frac{1}{2} \left( 1 - \left(\frac{b}{b_{\rm opt}}\right)^{2} \right) } \,,
\label{eq:delta-theta-y}
\end{equation}
where we have defined
\begin{equation}
b_{\rm opt} = \frac{1}{2} \sqrt{w_{0}^{2} + W_{0}^{2}}
\label{eq:delta-y_optimized}
\end{equation}
and
\begin{equation}
\langle \delta\theta_{y} \rangle_{\rm max} = A \, \frac{\mathcal{E}}{b_{\rm opt}^{3}} \,, \quad A = \frac{c\,n_{2,{\rm max}}}{4\pi \sqrt{e}} = 2.25 \, {\rm prad} \, \mu{\rm m}^{3} / {\rm J} \,.
\label{eq:max_deflection}
\end{equation}
$b_{\rm opt}$ is the value of the impact parameter $b$ at which $\langle \delta\theta_{y} \rangle$ is optimized. 
This optimal value is $\langle \delta\theta_{y} \rangle_{\rm max}$, the maximum possible barycenter shift given the pump energy and the pulse widths. 
For achievable pump energy $\mathcal{E}=2.5$~J with the LASERIX facility, and with a minimum waist at focus $w_0 = W_0 = 5 \ \mu$m for both the probe and the pump pulses, the average deflection angle is $\langle \delta\theta_{y} \rangle_{\rm max} = 0.13$~prad. 
Note that its functional form corresponds closely to that predicted in Eq.~(\ref{eq:delta-theta_scaling}) from purely dimensional considerations.  In this analysis, only the value $A$ is model-dependent; the value given here is for the Euler-Heisenberg model derived from QED (with orthogonal polarizations for the pump and probe).

An example of the mean deflection $\langle \delta\theta_{y} \rangle$ is shown, in solid black, in the left panel of Fig.~\ref{fig:delta-theta-y}, as a function of $b$ at fixed $\mathcal{E} = 2.5\,{\rm J}$ and $W_{0} = w_{0} = 5\,\mu{\rm m}$.  

Experimental constraints dictate that the tilt angle $\theta_{\rm tilt}$ cannot be exactly zero.  
Analytical calculations (see Appendix~\ref{Appendix-A}) show that the effect of the tilt angle is to introduce an overall correction factor $r_{\rm tilt}$ to the expression~(\ref{eq:delta-theta-y}) for the deflection angle: 
\begin{equation}
\langle \delta\theta_{y} \rangle = \langle \delta\theta_{y} \rangle_{\rm max} \, \frac{b}{b_{\rm opt}} \, e^{ \frac{1}{2} \left( 1 - \left(\frac{b}{b_{\rm opt}}\right)^{2} \right) } \times  r_{\rm tilt} \,,
\end{equation}
with
\begin{equation}
r_{\rm tilt} = \frac{{\rm cos}^{3}\left(\theta_{\rm tilt}/2\right)}{\sqrt{1 + \left(R^{2}-1\right) \, {\rm sin}^{2}\left(\theta_{\rm tilt}/2\right)}} \,, 
\quad R^{2} = \frac{w_{z}^{2}+W_{z}^{2}}{w_{0}^{2}+W_{0}^{2}} \,.
\label{eq:rtilt}
\end{equation}
Here, $W_{z}$ ($w_{z}$) is the longitudinal size of the pump (probe) pulse, defined analogously to the waist (i.e., the electric field amplitude is proportional to $e^{-z^{2}/w_{z}^{2}}$).
The parameter $R$ measures the ratio of the longitudinal to transverse sizes of the pulses. It is the only parameter that depends on their longitudinal size.  
To understand the form of the numerator of $r_{\rm tilt}$, consider two spherically symmetric pulses, so that $R = 1$ and the denominator is also just $1$.  In this case, the rotation of the pulse profiles is trivial, and the value of $r_{\rm tilt}$ comes from a combination of two factors: the refractive index change $\delta n$ is proportional to ${\rm cos}^{4}\left(\theta_{\rm tilt}/2\right)$ (see Eq.~(\ref{eq:n2})), while the interaction time during which the refractive index profile is able to act on the pulse is proportional to $1/{\rm cos}\left(\theta_{\rm tilt}/2\right)$.
Turning to values achievable with the LASERIX facility, taking a duration $T_{0} = 30\,{\rm fs}$, which corresponds to a longitudinal size $w_z = W_z = 7.6$~$\mu$m, and a minimum waist at focus $w_{0} = W_0 = 5\,\mu{\rm m}$, we find $R=1.5$.
Examples of $r_{\rm tilt}$, as a function of $R$ for several fixed values of $\theta_{\rm tilt}$, are shown in the right panel of Fig.~\ref{fig:delta-theta-y}.
Also shown in the left panel, in solid red,  is the theoretical prediction for the mean deflection $\langle \delta\theta_{y} \rangle$ as a function of $b$ with a tilt angle of $30\degree$ and a pulse duration (FWHM) of $30\,{\rm fs}$. For the optimal impact parameter $b_{\rm opt}$, the deflection is $\langle \delta\theta_{y} \rangle_{\rm max} = 0.11 \ \mathrm{prad}$, representing only a slight reduction with respect to the case of a ``head-on'' collision.  

As well as a deflection, the perturbed probe will also be characterized by a delay, on account of the slower wave speed due to the $\delta n$ induced by the pump.  For $\theta_{\rm tilt} = 0$, the delay accumulated at any point within the probe is $\delta t = \left(\int {\rm d}z \, \delta n\right)/c$, and is thus proportional to the bracketed expression in Eq.~(\ref{eq:Gaussian_theta_intensity}).  The mean delay $\langle \delta t \rangle$ is defined analogously to the mean deflection of Eq.~(\ref{eq:mean_deflection}), and is found to be:
\begin{equation}
\langle\delta t\rangle = \frac{n_{2,{\rm max}}}{4\pi} \, \frac{\mathcal{E}}{b_{\rm opt}^{2}} \, e^{-\frac{1}{2} \left(\frac{b}{b_{\rm opt}}\right)^{2}} \,.
\label{eq:mean_delay}
\end{equation}
This leads to a phase delay $\langle \delta\psi \rangle = \omega_{0}\,\langle \delta t \rangle$, where $\omega_{0}$ is the carrier frequency of the probe.  At a wavelength of $800\,{\rm nm}$, and for $b=b_{\rm opt}$ (so that the deflection $\langle \delta\theta_{y} \rangle$ is optimized), this yields an average phase delay of:
\begin{equation}
\langle \delta\psi \rangle = \left( 17.4 \, {\rm prad} \, \mu{\rm m}^{2} / {\rm J} \right) \times \frac{\mathcal{E}}{b_{\rm opt}^{2}} \,.
\label{eq:expected_phase_delay}
\end{equation}
Therefore, with $\mathcal{E}$ on the order of 1~J and $b_{\rm opt}$ on the order of a few $\mu{\rm m}$, the mean phase delay is on the order of a few picoradians, and is thus of the same order as the mean deflection angle.


\subsection{Amplification by the Sagnac interferometer}
\label{sub:amplification_sagnac}

In this section, we shall illustrate how the Sagnac interferometer yields an amplification of the barycenter shift in the measured intensity profile, considering a true {\it imperfect} interferometer.

\subsubsection{Description of scattering at a beamsplitter}
\label{subsub:scattering}
Let us consider a plane beamsplitter which is symmetrical with respect to translations and rotations within the plane.
Assuming it is lossless, the relation between the amplitudes of the incident and outgoing waves is given by a unitary scattering matrix, $S$:
\begin{equation}
\left[ \begin{array}{c} A_{1}^{\rm out} \\ A_{2}^{\rm out} \end{array} \right] = S \left[ \begin{array}{c} A_{1}^{\rm in} \\ A_{2}^{\rm in} \end{array} \right] \,.
\label{eq:S-matrix}
\end{equation}
The two waves with amplitudes $A_{1}^{\rm in}$ and $A_{2}^{\rm in}$ are incident from opposite sides with respect to the plane of the beamsplitter (see Fig.~\ref{fig:BS}), and similarly for the outgoing waves.

The most general $2 \times 2$ unitary matrix can be written in the following form:
\begin{equation}
S = e^{i \Phi_{0}} \left[ \begin{array}{cc} t \, e^{i \phi_{t}} & r \, e^{-i\phi_{r}} \\ -r \, e^{i\phi_{r}} & t \, e^{-i \phi_{t}} \end{array} \right] \,, \qquad t^{2} + r^{2} = 1 \,,
\label{eq:S-matrix}
\end{equation}
where the parameters $t$, $r$, $\phi_{t}$, $\phi_{r}$, and $\Phi_{0}$ are all real.  $t^{2}$ and $r^{2}$ are the fractions of the incident energy that are fed into the transmitted and reflected channels, respectively, and the losslessness of the beamsplitter is encoded in the unitarity relation $t^{2}+r^{2}=1$.  

The form of Eq.~(\ref{eq:S-matrix}) is very general, and thus independent of the internal structure of the beamsplitter.  However, it can prove useful to keep this internal structure in mind. It is shown explicitly in Fig.~\ref{fig:BS}.  The beamsplitter consists of a silica substrate, one surface being a 50/50 reflector/transmitter, the other having an anti-reflective coating so that, to a good approximation, rays are purely transmitted across it.  Each of the two surfaces separately can be described by a matrix of the form~(\ref{eq:S-matrix}).  For now we assume that, at the anti-reflective rear surface, $r_{\rm AR} = 0$; corrections in powers of $r_{\rm AR}$ will be included in Sec.~\ref{subsub:rAR_corrections} below, describing those rays which are reflected at this surface.

The meaning of the matrix $S$ is the following (and illustrated in Fig.~\ref{fig:BS}).  
When a ray is incident on the beamsplitter from one side, the transmitted and reflected rays pick up amplitudes $t$ and $r$, and phases $\Phi_{0}+\phi_{t}$ and $\Phi_{0}+\phi_{r}+\pi$, respectively.  When the ray is incident from the {\it other} side, the transmitted and reflected rays again pick up amplitudes $t$ and $r$, but now with phases $\Phi_{0}-\phi_{t}$ and $\Phi_{0}-\phi_{r}$, respectively~\footnote{Mathematically, the inclusion of a phase difference of $\pi$ for reflection from one side and not from the other simply accounts for the relative minus sign between the off-diagonal components of the $S$-matrix in Eq.~(\ref{eq:S-matrix}).  In the current context, it can be understood physically as being due to the fact that, from one side, there is reflection at an interface where the refractive index decreases, while from the other side, the reflection is at an interface where the index increases (see Fig.~\ref{fig:BS}).}.  The assumed translation and rotational symmetry of the beamsplitter means that these amplitudes and phases are independent of the point of incidence of the ray and of the direction (i.e., the azimuthal angle) from which it is incident.  We are mainly interested in beams which follow the same optical path, and since their interference will depend only on their {\it relative} phase, both the optical path length and the constant phase $\Phi_{0}$ 
drop out of the final result.  
It is for this reason that $\Phi_{0}$ is not included in the scattering amplitudes shown in Fig.~\ref{fig:BS}.

\subsubsection{Beamsplitter requirements for a Sagnac interferometer}

As discussed in Sec.~\ref{sec:setup}, the Sagnac interferometer makes use of  a single beamsplitter twice: first by splitting the incident probe pulse into two daughter pulses, which propagate around the same path in opposite directions; and then recombining these pulses at the output of the interferometer. This situation is described by a double application of the scattering matrix $S$.  The ideal Sagnac interferometer has complete destructive interference at the so-called ``dark port'' of the beamsplitter.  This imposes strict conditions on some of the parameters entering the scattering matrix: as shown in Fig.~\ref{fig:Sagnac_BS} (by applying the scattering amplitudes shown in Fig.~\ref{fig:BS}), the vanishing of the amplitude of the wave at the dark port requires $t^{2} e^{2 i \phi_{t}} - r^{2} = 0$ or, equivalently,
\begin{equation}
t^{2} = r^{2} = \frac{1}{2} \,, \qquad e^{2 i \phi_{t}} = 1 \,.
\end{equation}
The first condition is simply that the beamsplitter should be as close to 50/50 as possible, so that the two waves contributing to the output at the dark port have exactly the same amplitude.  The second condition is required to ensure that the two waves at the output are exactly out of phase, so that their interference is perfectly destructive.  Note that it is only $\phi_{t}$ that enters the second condition: $\Phi_{0}$ does not enter because it is only the relative phase between the two beams that is needed, while the fact that the reflected pulse reflects once from each side of the beamsplitter cancels out the phase $\phi_{r}$.

In reality, the beamsplitter will be imperfect: the transmission and reflection coefficients $t^2$ and $r^2$ will not be exactly $1/2$, and the phase $\phi_{t}$ will not be exactly null
(either because of an intrinsic asymmetry within the beamsplitter, or because of other systematics that affect the relative optical path length).
We parameterize the difference by introducing the small quantities $\delta a$ and $\delta\phi$:
\begin{eqnarray}
& t^{2}  =  \frac{1}{2} \left(1+\delta a\right) \,, \qquad r^{2}  =  \frac{1}{2} \left(1-\delta a\right) \,, \nonumber \\
& e^{2i\phi_{t}}  =  e^{2i \,\delta\phi} \approx 1 + 2 i \, \delta\phi \,.
\label{eq:imperfect-beamsplitter}
\end{eqnarray}
Then, neglecting terms of second order and higher in $\delta a$ and $\delta\phi$, the amplitude $A_{\rm dark}$ of the outgoing wave at the dark port, relative to the incident amplitude $A_0$ entering inside the interferometer, is
\begin{equation}
A_{\rm dark}/A_0 = t^{2}e^{2i\phi_{t}} - r^{2} \approx \delta a + i \, \delta\phi \,,
\label{eq:amplitude-dark-output}
\end{equation}
The extinction factor $\mathcal{F}$, defined as the ratio of the outgoing intensity at the dark port to the input intensity, is then
\begin{equation}
\mathcal{F} = \left(\delta a\right)^{2} + \left(\delta\phi\right)^{2} \,.
\label{eq:extinction-factor}
\end{equation}

\subsubsection{Intensity profile of the interference in the dark output: amplification of the barycenter shift}

At focus in the interaction area, the probe field is refracted by the pump through an average deflection angle $\langle \delta\theta_{y} \rangle$ given by Eq.~(\ref{eq:delta-theta-y}), and is delayed by an average phase delay $\langle \delta\psi \rangle$ given by Eq.~(\ref{eq:expected_phase_delay}). After recollimation, the deflection angle becomes a transverse displacement of the probe field, whose magnitude depends also on the focal length $f$ of the lenses used to focus the probe.  Assuming $b = b_{\rm opt}$, we have simply
\begin{equation}
 \langle \delta y \rangle_{\rm direct} = f \times \langle \delta\theta_{y} \rangle_{\rm max} \,.
\label{eq:theoretical-signal}
\end{equation}
We call this the ``direct'' (or ``bare'') deflection, since it is the shift in the barycenter of the intensity that would be measured if the probe were detected directly, without interfering with the unperturbed probe.  
As will be shown now, the interference of the probe pulse with the unperturbed reference pulse in the dark output of the Sagnac interferometer allows this transverse displacement to be amplified.

We denote by ${\bf E_{\rm pr}}$ and ${\bf E_{\rm re}}$ the slowly-varying envelopes of, respectively, the perturbed probe electric field, and the unperturbed reference electric field, in the dark output of the Sagnac interferometer.  The full electric fields include a rapidly oscillating carrier wave:
${\bf E}_{\rm full} = {\rm Re}\left\{{\bf E} \, e^{i\left(k_{0}z - \omega_{0}t\right)}\right\}$. 
Denoting by ${\bf E_{\rm 0}}$ the incident field entering the interferometer, we have 
\begin{eqnarray}
{ E}_{\rm re}(x,y) &=& t^{2}\ e^{2i\delta \phi} { E}_{\rm 0}(x,y) \,, \\
{ E}_{\rm pr}(x,y)&=& r^{2}\ e^{-i\delta \psi}{ E}_{\rm 0} (x,y-\langle \delta y \rangle_{\rm direct}) \,,
\nonumber
\label{eq:fields}
\end{eqnarray}
where we have approximated the perturbation as being fully characterized by $\langle\delta y\rangle$ and $\langle\delta\psi\rangle$ (so that the profile of the field is otherwise unaffected).
In the dark output, the two fields interfere destructively and the resultant field is
\begin{eqnarray}
{ E}_{\rm dark} ={ E}_{\rm re}-{ E}_{\rm pr} \,,
\label{eq:darkE}
\end{eqnarray}
with corresponding intensity
\begin{eqnarray}
I_{\rm dark} &=& \frac{c}{2} \epsilon_{0} \, \left|E_{\rm dark}\right|^{2} \,. 
\end{eqnarray}
The data are images obtained after integrating the intensity over the pulse duration.
Since both pulses cross the same amount of material in the beamsplitter, they are equally distorted by dispersive effects, so the integrated intensity contains all relevant information.
Up to negligible higher order terms, $I_{\rm dark}$ reads
\begin{eqnarray}
\label{eq:I_dark0}
I_{\rm dark} &=& \frac{c}{8}\epsilon_0\left\vert \left(1+{\delta a}\right)(1+2i\delta \phi) E_{\rm 0}(x,y)\right. \\
&-& \left. \left(1-{\delta a}\right)\left(1-i\langle \delta \psi \rangle\right)
E_{\rm 0}(x,y-\langle \delta y \rangle_{\rm direct})\right\vert^2\nonumber
\end{eqnarray}
which gives, after reordering,
\begin{eqnarray}
\label{eq:I_darkEnd}
I_{\rm dark} &=& \delta a^2\  I_0\left(x,y+\frac{\langle \delta y \rangle_{\rm direct}}{2\delta a}\right) \\
&& \qquad + \delta \phi^2\  I_0(x,y)\left(1+\frac{\langle \delta \psi \rangle}{\delta \phi}\right) \,, \nonumber
\end{eqnarray}
where $I_0(x,y)  = \frac{1}{2} c \epsilon_{0} E_{\rm 0}^{2}(x,y)$ 
is the intensity profile of the incident pulse.

Equation~(\ref{eq:I_darkEnd}) is the generalization of Eq.~(\ref{eq:extinction-factor}) in the presence of a pump crossing effect.
The measured intensity profile in the dark output depends linearly on both the average phase delay and the average deflection induced by the pump pulse on the probe pulse.
The direct lateral shift $\langle \delta y \rangle_{\rm direct}$ is amplified by a factor of $1/(2\delta a)$, but $\delta \phi$
adds an unshifted intensity component. 
Since $\langle \delta \psi \rangle \ll \delta \phi$, the combination of both components shifts the barycenter of the output signal by
\begin{equation}
\langle \delta y \rangle_{\rm Sagnac} = \frac{\delta a/2}{(\delta a)^2 + (\delta\phi)^2} \langle \delta y \rangle_{\rm direct} = \frac{\delta a}{2\mathcal{F}} \langle \delta y \rangle_{\rm direct} \,.
\label{eq:direct_v_Sagnac}
\end{equation}
 It is clear that being able to extract the vacuum index perturbation from the output intensity profile requires knowledge not just of the extinction factor $\mathcal{F}=\left(\delta a\right)^{2} + \left(\delta\phi\right)^{2}$ but also of the values of $\delta a$ and $\delta\phi$ separately.

The main contribution to $\delta\phi$ comes from understood optical defects; these will be explained in more detail in Section \ref{sec:extinction}. 
There are two other expected physical contributions to $\delta\phi$:
the phase asymmetry intrinsic to the beamsplitter, and the Sagnac phase shift arising from the rotation of the Earth, on the order of $3\times10^{-6}$ rad for a 1 metre Sagnac arm length.  Both of these additional contributions turns out to be negligible with respect to the reflection/transmission asymmetry $\delta a$ of the beamsplitter, which in the DeLLight prototype is on the order of $10^{-3}$ (see Section~\ref{sec:extinction}).

The target specification of the final design is to achieve an extinction factor $\mathcal{F}$ consistently dominated by $\delta a$, i.e., with negligible phase asymmetry $\delta \phi$ compared to $\delta a$. In this case, Eq.~(\ref{eq:direct_v_Sagnac}) becomes
\begin{equation}
\langle \delta y \rangle_{\rm Sagnac} = \frac{1}{2 \, \delta a} \langle \delta y \rangle_{\rm direct} = \frac{1}{2\sqrt{\mathcal{F}}} \langle \delta y \rangle_{\rm direct} 
\label{eq:direct_v_Sagnac-2}
\end{equation}
and the amplification factor multiplying the lateral shift is simply equal to $1/(2\sqrt{\mathcal{F}})$.

We also note that the second term in (\ref{eq:I_darkEnd}) gives access to another way to search for  
a vacuum index modification, since it is sensitive to $\langle \delta \psi \rangle$ thanks to a linear effect on the total intensity. In this case $1/\delta \phi$ amplifies the sensitivity. In principle, one might optimize a setup aiming at a $\langle \delta \psi \rangle$ measurement, but sensitivity estimates show this is out of reach of the LASERIX facility.

\subsubsection{Corrections due to reflection at the anti-reflective surface
\label{subsub:rAR_corrections}}

We now consider corrections due to rays which reflect from the anti-reflective rear surface of the beamsplitter as illustrated in Figure~\ref{fig:back-reflexions}.
The rays which reflect once from this surface do not overlap with the main signal described above; instead, they are separated from the main signal by a distance that depends on the thickness of the substrate of the beamsplitter.  
These {\it back-reflections} are useful in following, and correcting for, the beam pointing fluctuations of the laser beams; this is described in Sec.~\ref{subsec:beampointing} below.  
However, when considering rays that reflect twice from the anti-reflective rear surface, there is one such ray that overlaps both spatially and temporally with the main signal (in the sense that it follows the same optical path length when traversing the interferometer), with an amplitude at the dark port equal to $-t^2\,r_{\rm AR}^{2} \approx -r_{\rm AR}^{2}/2$.
The total amplitude at the dark port is therefore given by (neglecting terms in $r_{\rm AR}^{2}\delta a$ and  $r_{\rm AR}^{2}\delta\phi$)
\begin{equation}
A_{\rm dark}/A_{0} \approx \delta a - \frac{1}{2} r_{\rm AR}^{2} + i \, \delta\phi \,,
\label{eq:amplitude_rAR}
\end{equation}
and the extinction factor is now
\begin{equation}
\mathcal{F} \approx \left( \delta a - \frac{1}{2} r_{\rm AR}^{2} \right)^{2} + \left(\delta\phi\right)^{2} \,.
\label{eq:extinction_rAR}
\end{equation}
Therefore, an anti-reflectivity coating with $r_{\rm AR}^{2} \ll \delta a$ is required to avoid any contribution of the back-reflections to the extinction factor of the interferometer.

\subsection{Expected sensitivity}
\label{sec:sensitivity}

To account for inevitable low-frequency variations of the measured position of the unperturbed probe, we propose to alternate shots with and without interaction between the pump and probe pulses, which we respectively label as ``ON'' and ``OFF'' measurements.
We extract the barycenters of the intensity profile measured in the dark output of the interferometer for successive ON and OFF measurements, which we name $\bar{y}^{\rm ON}(i)$ and $\bar{y}^{\rm OFF}(i)$, respectively.  The signal $\delta y (i)$ of the $i^{\rm th}$ ``ON-OFF'' measurement, corresponding to a shift of the barycenter due to the interaction with the pump, is then defined as
\begin{eqnarray}
\delta y (i) = \bar{y}^{\rm ON}(i) - \bar{y}^{\rm OFF}(i) \,.
\end{eqnarray}
The measured $\delta y (i)$ will have a certain distribution characterized by its mean value $\langle{\delta y}\rangle$ and its standard deviation $\sigma_{y}$ (which we shall henceforth refer to as the {\it spatial resolution}).
Collecting $N$ such measurements, the statistical error (one standard deviation) of the observed mean 
$\langle{\delta y}\rangle$ is 
equal to $\sigma_y/\sqrt{N}$. 

Since classical electromagnetism predicts no light-by-light refraction (and hence $\delta y = 0$), we wish to measure a signal whose difference from zero is statistically significant, i.e., for which the measured mean $\langle \delta y \rangle$ is a few standard deviations $\sigma_{y}/\sqrt{N}$ away from zero. 
Assuming that the average signal $\langle{\delta y}\rangle$ is equal to $\langle \delta y \rangle_{\rm Sagnac} = \langle \delta y \rangle_{\rm direct}/\left(2 \sqrt{\mathcal{F}} \right)$ of Eq.~(\ref{eq:direct_v_Sagnac}), 
the sensitivity of the experiment, in terms of the number of standard deviations $N_{\mathrm{sd}}$, is
\begin{eqnarray}
\label{eq:Nsd-sensitivity-1}
N_{\mathrm{sd}} = \frac{\langle \delta y \rangle_{\mathrm{Sagnac}}}{\sigma_y/\sqrt{N}} = \frac{\langle \delta y \rangle_{\mathrm{direct}}/\left(2 \sqrt{\mathcal{F}}\right)}{\sigma_y/\sqrt{N}} \,.
\end{eqnarray}
Using Eqs.~(\ref{eq:delta-y_optimized}), (\ref{eq:max_deflection}) and~(\ref{eq:theoretical-signal}), 
the number of standard deviation of the signal, in terms of the experimental parameters, is
\begin{eqnarray}
\label{eq:Nsd-sensitivity-2}
N_{\mathrm{sd}} =  \frac{c \, n_{2,\mathrm{max}}}{\pi \sqrt{e}} \times \frac{\mathcal{E} \times f \times r_{\rm tilt}\left(\theta_{\rm tilt}\right) \times \sqrt{N}}{(w_0^2 + W_0^2)^{3/2} \times \sqrt{\mathcal{F}} \times \sigma_y} \, ,
\end{eqnarray}
where $f$ is the focal length of the lenses used to focus the probe and reference pulses inside the Sagnac interferometer, and $r_{\rm tilt}$ is the correction factor defined in Eq.~(\ref{eq:rtilt}).
The number of  ``ON-OFF'' measurements is related to the repetition rate $f_{\mathrm{rep}}$ of the laser shots and to the the total duration of the experiment, $T_{\mathrm{obs}}(\mathrm{days})$, given in number of days: $N = 86400 \times f_{\mathrm{rep}} ({\rm Hz}) \times T_{\mathrm{obs}}(\mathrm{days})$. 
With experimental parameters given in local units, and for the expected QED signal ($n_{2,\mathrm{max}}=n_{2,\mathrm{QED}}$), we get
\begin{eqnarray}
\label{eq:Nsd-sensitivity-3}
N_{\mathrm{sd}}  &=& 0.6 \times \sqrt{f_{\mathrm{rep}}({\rm Hz})}  \times \mathcal{E}(\mathrm{J}) \times \sqrt{T_{\mathrm{obs}}(\mathrm{days})} \times  \\ 
&& \quad  \frac{f(\mathrm{mm})   \times r_{\rm tilt}\left(\theta_{\rm tilt}\right) }{(w_0^2(\mu \mathrm{m}) + W_0^2(\mu \mathrm{m}))^{3/2} \times \sqrt{\mathcal{F}/10^{-5}} \times \sigma_y(\mathrm{nm})} \,. \nonumber
\end{eqnarray}
As detailed in Sec.~\ref{sec:prototypes}, preliminary experimental tests show that an extinction factor of the Sagnac interferometer $\mathcal{F} = 0.4 \times 10^{-5}$ (corresponding to an amplification factor $1/\left(2 \sqrt{\mathcal{F}}\right)$ = 250) and a spatial resolution $\sigma_y = 10~{\rm nm}$ can be achieved.  A waist at focus $W_0=5 \ \mathrm{\mu}$m for the pump beam is a standard value reachable with intense pulses.  A same waist at focus $w_0=5 \ \mathrm{\mu}$m for the probe beam can be obtained by requiring a waist $w \simeq 25~{\rm mm}$ for the collimated probe beam in the interferometer (before focus) and by using optical lenses with focal lengths $f = 500$~mm.  A conservative tilt angle of $30\degree$ yields to a correction factor $r_{\rm tilt} \approx 0.9$.
With an energy of the pump pulse of 2.5~J as delivered by the LASERIX facility, the expected signal in the dark output of the interferometer\footnote{The direct signal, using a direct pointing method at a distance $f=500$~mm without interferometer, would be only $\langle \delta y \rangle_{\rm direct}= 60$~fm.} is $\langle \delta y \rangle_{\rm Sagnac} = 15$~pm. With the relatively high repetition rate of LASERIX $f_{\mathrm{rep}}=10$~Hz, we get $N_{\mathrm{sd}} = 0.95 \times \sqrt{T_{\mathrm{obs}}(\mathrm{days})}$, which means that the sensitivity of the expected QED signal at 1 standard deviation (1-sigma) is obtained after 1.1 days of collected data, or a 5-sigma observation after about one month.

\subsection{Contribution of residual gas}

The optical Kerr effect in the residual gas is at first sight a nonlinear effect much stronger than the corresponding 
effect in vacuum. For instance, the Kerr index in air at atmospheric pressure ($n_2 \simeq 3 \times 10^{-19} \, {\rm cm}^{2}/{\rm W}$) is $2 \times 10^{14}$ times larger than the expected nonlinear index $n_{2,{\rm QED}}$ of vacuum. 
Since this value is proportional to the pressure, setting $n_{2}$ to lie an order of magnitude below $n_{2,{\rm QED}}$ requires a residual pressure of $\sim 10^{-12}$~mbar.
However this results is true only if the  pump-probe interaction volume $V$ is large enough to contain a few atoms.
For the DeLLight experiment, the interaction volume is very small, of the order of $V \simeq w_0^2 \times \Delta t \times c \simeq 225 \ \mu \mathrm{m}^3$. 
The vacuum chamber of the DeLLight experiment is designed to ensure a nominal pressure for the full chamber below $10^{-6} $~mbar and a local pressure in the interaction area below $10^{-8} $~mbar. 
At this local pressure, there is on average only 0.1 residual atom inside the pump-probe interaction volume $V$.
Therefore, at achievable pressures, we are already in a regime far below that of the coherence required for a possible refraction effect, and where the notion of refractive index for air can no longer be defined.

Moreover, in our setup using the LASERIX facility, the intensity of the pump in the 
interaction area is of the order of $10^{20}$~W/cm$^2$. At this intensity, the residual gas is completely 
ionized by the pump, all the electrons being removed from the atoms. 
We thus have an electromagnetic wave crossing a pure relativistic plasma. 
Investigating the dynamics of this system will require numerical simulations, but its effect on the DeLLight signal is expected to be negligible. 
Indeed the plasma density and the plasma index is transversally uniform in the pump-probe interaction area, so that the perturbation of the probe phase is symmetric and the mean deflection (the signal) is zero

Experimental tests are feasible that could distinguish between a possible artefact induced by the residual gas and the vacuum signal. 
First, by decreasing the intensity of the pump by a factor of 10, the vacuum signal 
should be reduced by the same factor, while any artefact due to the plasma 
should be constant (the intensity still being high enough to completely ionize the atoms). 
Also, by inverting the propagation directions of the probe and reference pulses (i.e., moving from a counter-propagating pump-probe test to a co-propagating test), any plasma signal should remain while the vacuum signal must be suppressed according to the $r_{\rm tilt}$ factor of Eq.~(\ref{eq:rtilt}). 

Finally we add that the Kerr effect in air will be used to calibrate and validate the DeLLight method. The measurement will be carried out with a relatively low intensity ($\sim10^{11}$~W/cm$^2$) in order to avoid generating a plasma via ionization of air molecules, and the Kerr effect will be measured as a function of the pressure. 
More generally, the use of the Kerr or plasma signals can be used to monitor and control the spatial and temporal overlap of the pump and probe pulses in the interaction area.

\section{Preliminary results of the DeLLight prototype
\label{sec:prototypes}}
A prototype of the experiment has been developed in order to study and measure two critical experimental parameters: the extinction factor of the Sagnac interferometer $\mathcal{F}$ and the spatial resolution $\sigma_y$.

\subsection{Description of the prototype}

Two different Sagnac interferometers have been developed and tested: first, a rectangular Sagnac interferometer composed of three mirrors, without any focus inside the interferometer; then a triangular configuration 
composed of two mirrors and two optical lenses so as to 
focus the laser pulses at the midpoint 
of the interferometer. Both geometries are insensitive to beam pointing fluctuations in angle and transverse position.

The optical setups are shown in Figure~\ref{fig:setup-prototype}. 
The incident laser pulses are delivered by the LASERIX beam with a repetition rate of 10 Hz. The  energy per pulse is about 20 $\mu$J, and the pulse duration is 50~fs. The central wavelength is $\lambda_0 = 815$~nm with a spectral width $\Delta \lambda = 40$~nm (FWHM). 
A spatial filter composed of two spherical lenses and a pinhole located in the focal point of the telescope produces a smooth transverse intensity profile that is close to Gaussian. 
An initial beamsplitter (BS-1) provides two distinct beams: the 
probe beam, used here only to study the performance of the 
interferometer, and the 
pump beam which will be used in a second step to validate the DeLLight technique by measuring the Kerr effect in a 
medium. 
For the measurements presented in this article, the pump beam is stopped. Results of the measurements of the Kerr effect in a medium 
will be presented in a future article. 

Before entering inside the interferometer, a polarizer selects the horizontally polarized (p-pol.) component and a neutral density filter sets 
the suitable intensity of the incident probe pulse.
The width $w$ of the intensity profile of the probe pulse is $w \approx 1$~mm (FWHM).

The beamsplitter BS-2 of the Sagnac interferometer is a 50/50 commercial femtosecond p-pol beamsplitter (Semrock FS01-BSTiS-5050P-25.5). Its coating is produced by the Ion Beam Sputtering technique which delivers uniform atomic layers. The expected theoretical values of the reflection and transmission coefficients $r^2$ and $t^2$ vary in a range $r^2 = t^2 = 50 \pm 0.2 \%$ over a broad spectrum between 650 and 1100~nm. 
Using Eq.~(\ref{eq:extinction-factor}), a deviation of $0.2\%$ (equivalent to $\delta a = 4\times 10^{-3}$) corresponds to an extinction factor $\mathcal{F} = (\delta a)^2 = 1.6 \times 10^{-5}$. The phase factor $\phi$ of the beamsplitter (Eq.~(\ref{eq:imperfect-beamsplitter})) is not characterized by the producer. 
The thickness of the beamsplitter is 3~mm and the group delay dispersion (GDD) is less than 30~fs$^2$ for both reflection and transmission. An anti-reflection coating has been deposited on the rear side of the beamsplitter with a reflectivity coefficient $r_{\mathrm{AR}}^2= 0.1 \ \%$ at 800~nm. 
The mirrors inside the interferometer are standard femtosecond dielectric mirrors with a low GDD value (typically less than 50~fs$^2$) and standard laser grade surface qualities: a  flatness peak-to-valley $<\lambda/10$ at 633~nm, a quality 20-10 Scratch-Dig, and a  roughness with typical $\mathrm{RMS} <5 \ \mathring{\mathrm{A}}$.
The beamsplitter and one of the mirrors are controlled by kinematic mirror mounts with two static piezoelectric adjusters for horizontal and vertical alignment with an angular resolution of $0.5 \ \mu$rad for a 0.1~V step. 
The lateral position of one mirror is controlled by a micrometric translation stage. 
For the triangular configuration of the Sagnac interferometer, the beam is focused using two best form optical lenses L-1 and L-2, placed between the two mirrors M-1 and M-2.  The lenses have equal focal length $f=100$~mm, and are separated from each other by a distance $2f$. 
One lens is mounted on a 3-axis piezoelectric adjuster.
The length of the longest interferometer arm between the two mirrors M1 and M2 is about 40~cm.

The dark output of the interferometer is read by a CCD camera (Basler acA1300-60gm) containing $1024 \times 1280$ pixels. The pixel dimension is $5.3 \times 5.3 \ \mathrm{\mu m}^2$ and the maximum charge storage capacity before saturation (full well capacity) is $10^4$ electrons per pixel. 
An interferential multilayer dielectric filter $\Delta\lambda=3$~nm centered at 808~nm is placed 
in front of the CCD camera. Rotation of the incident angle of the filter allows us to select a wavelength from 808 to 800~nm and thereby 
optimize the extinction factor by minimizing the deviation coefficient $\delta a$ 
of the beamsplitter. Since $r^2$ and $t^2$ depend also on the polarization of the incident beam, the deviation coefficient $\delta a$ is also minimized by rotating the incident polarization with a half-wave plate WP installed just after the polarizer, and before the Sagnac interferometer.

\subsection{Extinction factor}
\label{sec:extinction}

The extinction has been measured first for the rectangular Sagnac interferometer with three mirrors and no focusing. 
A typical transverse intensity profile recorded by the CCD camera in the dark output of the interferometer, after optimization of the extinction factor,  is presented in Fig.~\ref{fig:extinction-max}.
The two 
spots due to 
back-reflections on the rear side of the beamsplitter are clearly observed on opposite lateral sides of the main signal, with approximately equal intensities $I_{\rm AR}$.  
As illustrated in Fig.~\ref{fig:back-reflexions},~\footnote{While Fig.~\ref{fig:back-reflexions} shows the case of a triangular two-mirror geometry with 
focusing, the contributions of the back-reflected beams to the output intensity are the same as in the case of a rectangular three-mirror geometry without focusing.} one spot $I_{\rm AR,1}$ corresponds to the direct image of the incident intensity after a single reflection on the rear side of the beamsplitter, while the second spot $I_{\rm AR,2}$ is a superposition  
of four distinct reflected waves : 
\begin{eqnarray}
I_{\rm AR,1} &=& I_{\rm in} \times r_{\rm AR}^{2} \, r^{2} \nonumber \\
I_{\rm AR,2} &=& I_{\rm in} \times r_{\rm AR}^{2} \, r^{2} \left| 1 + 2 \left( t^{2} e^{2i \delta\phi} - r^{2} \right) \right|^{2} \,.
\end{eqnarray}
where $I_{\rm in}$ is the incident intensity 
at the entrance of the Sagnac interferometer. 
At lowest order, 
we have $r^{2} = t^{2} = 1/2$ and $\delta\phi = 0$, which gives $I_{\rm AR,1} = I_{\rm AR,2} = I_{\rm in} \times r_{\rm AR}^2 /2$.
The attenuation factor $\mathcal{F}_{\rm AR}$ of the back-reflections is $\mathcal{F}_{\rm AR}=I_{\rm AR}/I_0 = r^2_{\rm AR} / 2$. It has been measured with a photodiode and a set of calibrated neutral density filters, and is $ \mathcal{F}_{\rm AR} = 4 \times 10^{-4}$. It corresponds to $r^2_{\rm AR} = 8 \times 10^{-4}$ at 800~nm, in agreement with the typical value given by the producer. 
For an incidence angle of $45\degree$, the transverse distance between the back reflections and the interference signal is $\delta_{\rm AR} = e/\sqrt{n^2-1/2}$, 
where $e$ is the thickness of the beamsplitter and $n$ the refractive index of its substrate. Here, $e=3$~mm, $n \simeq 1.45$, and $\delta_{\rm AR} \simeq 2.37$~mm.

The interference signal is located in the central part, delimited by the dotted white circle in Fig.~\ref{fig:extinction-max}. 
The residual phase noise of the interference signal is shown in the right image of Fig.~\ref{fig:extinction-max} by increasing the sensitivity scale of the display (the intensity of the back-reflections cannot be directly attenuated using a neutral density filter because of the too small thickness of the beamsplitter and therefore a too small transverse distance between the signal area and the back-reflections).  
Two types of noise pattern can be distinguished: hot spots in the central area of the expected intensity signal and interference rings with large transverse size (low spatial frequency).  
The extinction factor measured in the central area is at worst $ \mathcal{F} \simeq 2 \times 10^{-5}$. 

The noise pattern is due to a difference in phase between the two counter-rotating pulses when they interfere; we thus refer to it as the {\it phase noise pattern}. 
It is induced by surface defects on the mirrors and on the substrate of the beamsplitter. 
From Eq.~(\ref{eq:extinction-factor}), the observed phase noise at a level $\mathcal{F}  = 2 \times 10^{-5}$ corresponds to a difference of phase  $\delta \phi = 4$~mrad. It is equivalent to a difference in the optical path lengths $\delta l = \delta \phi \times \lambda_0 /(2\pi) \simeq 5 \ \mathring{\mathrm{A}}$. This value is in agreement with the current tolerance of the surface quality of the mirrors and beamsplitter, indicating that the residual hot spots seems to be produced by roughness defects on the surface of the mirrors or the substrate of the beamsplitter. This is experimentally confirmed by the fact that the pattern of the noisy hot spots is horizontally translated when the position of the incident beam is slightly horizontally translated (by the translation stage TS-1 shown in Fig.~\ref{fig:setup-prototype}), allowing to scan different part of the surface of the mirrors and beamsplitter. 
The probable origin of the interference rings, spread over a large transverse size, is the relatively limited quality of the flatness of the mirrors and the beamsplitter, and a modest parallelism of the beamspitter ($< 5$~arcminutes).
Another possible origin of the residual noise is a deviation from a symmetric $50/50$ beamsplitter. However the dielectric coating is expected to be very uniform at the atomic scale. Therefore the asymmetry should be also uniform and should not exhibit local hot spots.

An additional limitation of the extinction is the presence of a double back-reflection which interferes with the interference signal, as discussed in Sec.~\ref{subsub:rAR_corrections} (see Eqs.~(\ref{eq:amplitude_rAR}) and~(\ref{eq:extinction_rAR})).
For the beamsplitter used in the prototype, we have $\delta a \leq 4\times 10^{-3}$ and $r^2_{\rm AR} \simeq 10^{-3}$.
Therefore, the contribution from the additional term is small.

In order to reduce the phase noise, and thereby reach an extinction factor on the order of $10^{-6}$, a new beamsplitter and mirrors are under development with superpolished surface qualities: surface flatness peak-to-valley $<\lambda/100$ at 633~nm, surface roughness $\mathrm{RMS} < 1 \ \mathring{\mathrm{A}}$, parallelism of the beamsplitter $< 1$~arcsecond, and an anti-reflection coating on the rear side of the beamsplitter with a reflectivity coefficient $r^2_{\mathrm{AR}} \sim 10^{-4}$ at 800~nm. 

The extinction has also been measured for a triangular Sagnac interferometer with a focusing of the probe beam by two optical lenses. 
The minimum waist measured at focus is $w_0 \approx 25 \ \mu$m.  
A similar quality of the global extinction has been measured in the dark output of the interferometer. Degradation of the extinction by a factor about 2 can be observed locally due to additional defects of lens surfaces.
Focusing with off-axis parabolic mirrors (rather than with lenses) is also envisaged.

\subsection{Spatial resolution}
\label{sec:spatial-resolution}

\subsubsection{Shot noise}

The spatial resolution for the measurement of the barycenter 
of the intensity profile is inherently limited by the intrinsic shot noise related to the statistical fluctuations of the average number of photoelectrons $\sqrt{N_{\rm p.e.}}$ detected by the CCD camera. The spatial resolution scales as $w/\sqrt{N_{\rm p.e.}}$ where $w$ is the beam width on the CCD camera. If $n_{\rm p.e.}$ is the average number of photoelectrons per pixel, and $d_{\rm pix}$ is the side length of each square pixel, then the spatial resolution scales as $d_{\rm pix}/\sqrt{n_{\rm p.e.}}$ and is independent of the beam width. Therefore, in order to achieve the best spatial resolution, we need the highest charge storage capacity before saturation per unit surface of the optical readout. For a CCD camera, this is referred to as the {\it full well capacity} $N_c$, which corresponds to the maximum number of stored electrons per pixel before saturation.

The spatial resolution has been calculated using Monte Carlo simulations with a transverse Gaussian intensity profile and an average number of electrons per pixel at maximum equal to 75$\%$ of the full well capacity $N_c$ in order to ensure a linear response of the pixels (as is the case in the experimental data). Accounting for both the corrections of 
beam pointing fluctuations (see below) 
and the ``ON-OFF'' subtraction procedure defined in section~\ref{sec:sensitivity}, the ultimate spatial resolution is found to be:
\begin{equation}
\sigma_y(\mathrm{shot \ noise}) = 0.75 \times \frac{d_{\rm pix}}{\sqrt{N_c}}
\label{eq:shot-noise}
\end{equation}
We also verify that the dark current contribution is negligible with respect to the shot noise. 
With the standard Basler CCD camera currently used in the prototype ($d_{\rm pix} = 5.3 \ \mu$m and  $N_c = 10^4$ electrons per pixel), the expected spatial resolution limited by the shot noise is 
$\sigma_y(\mathrm{shot \ noise}) \approx 40$~nm.
With the new generation of standard CCD cameras (for instance $d_{\rm pix} = 1.8 \ \mu$m and $N_c = 10^4$ electrons per pixel for the Basler acA4024-29um camera), a spatial resolution 
$\sigma_y(\mathrm{shot \ noise}) \approx 13$~nm
can be achieved. Resolution down to 10~nm can be reached with CCD cameras with larger full well capacity per unit surface, or with the direct use of position sensitive photodiodes.

\subsubsection{Beam pointing fluctuations}
\label{subsec:beampointing}

In the current prototype, no system of laser beam stabilization has been installed.  Significant beam pointing fluctuations are present, leading to large fluctuations 
of the transverse position of the intensity profile measured by the CCD.  
These are characterized by a standard deviation on the order of $\sigma_b  = 10-15 \ \mu$m, depending on the experimental conditions in the laser room.  
The online monitoring of the beam position is performed using the back-reflection which corresponds to the direct image of the incident beam (see Figure~\ref{fig:back-reflexions}).
Ideally, in order to maintain a high extinction, the intensity of the back-reflections must be attenuated with a neutral density so as to be of the same order as the interference signal.  
However, the back-reflected spots are not sufficiently distant 
from the interference signal because the thickness of the beamsplitter is too small
(the thickness of the future beamsplitters under development will be increased to $e=6.35$~mm for this purpose). 
Therefore, we have had to reduce the extinction of the interferometer in order to obtain an interference signal intensity equal to the back-reflection intensities.
This was done by rotating very slightly the incident polarization of the laser pulse with the half-wave plate WP (see Figure~\ref{fig:setup-prototype}), increasing the deviation coefficient $\delta a$ 
of the beamsplitter.  Figure~\ref{fig:rotated-polar} shows the intensity profile recorded by the CCD after rotation of the polarization with the interference signal intensity equal to the back-reflection intensity. The measurement of the spatial resolution is performed in this configuration. It corresponds to an extinction factor $\mathcal{F} = 5 \times 10^{-4}$.

We present here the measurement of the spatial resolution, obtained with 8000 successive laser shots (at a 10~Hz repetition rate), 
collected with the rectangular Sagnac interferometer with three mirrors and no focusing.
Data of successive odd $(2i-1)$ and even $(2i)$ laser shots are arbitrarily separated into OFF and ON data in order to define an ``ON-OFF'' measurement $i$ using two successive laser shots (at a 5~Hz repetition rate) . 
The barycenters of the intensity profiles of the interference signal $\bar{y}_{\mathrm{sig}}(i)$ and the back-reflection $\bar{y}_{\mathrm{ref}}(i)$ are calculated along the vertical 
axis, using a square analysis window (or {\it Region of Interest}) whose size $w_{\rm RoI}$ is by default equal to half the width $w$ (FWHM) of the intensity profile. 
The beam pointing fluctuations are suppressed for each ON and OFF measurement using the correlation of the barycenters of the signal $\bar{y}_{\mathrm{sig}}(i)$ and the back-reflection $\bar{y}_{\mathrm{ref}}(i)$. One obtains the corrected positions:
\begin{eqnarray}
\bar{y}_{\mathrm{corr}}^{\mathrm{OFF}}(i) & = & \bar{y}_{\mathrm{sig}}^{\mathrm{OFF}}(i) - \left( a_{\mathrm{OFF}} \times \bar{y}_{\mathrm{ref}}^{\mathrm{OFF}}(i) + b_{\mathrm{OFF}} \right) \nonumber \\
\bar{y}_{\mathrm{corr}}^{\mathrm{ON}}(i) & = & \bar{y}_{\mathrm{sig}}^{\mathrm{ON}}(i) - \left( a_{\mathrm{OFF}} \times \bar{y}_{\mathrm{ref}}^{\mathrm{ON}}(i) + b_{\mathrm{OFF}} \right) 
\label{eq:BPcorrections}
\end{eqnarray}
where $a_{\mathrm{OFF}}$ and $b_{\mathrm{OFF}}$ are obtained by fitting the linear correlation, using only the OFF measurements. 
Figure~\ref{fig:beam-pointing-corrections} shows the correlation for both OFF and ON data, with the result of the linear fit superposed. 
The signal $\delta y (i)$ of the ``ON-OFF'' measurement $i$  is
\begin{eqnarray}
\delta y (i) = \bar{y}_{\mathrm{corr}}^{\mathrm{ON}}(i) - \bar{y}_{\mathrm{corr}}^{\mathrm{OFF}}(i).
\end{eqnarray}
Its average value $\langle \delta y \rangle$ is expected to be zero since there is no interaction between pump and probe pulses.
The distribution of the raw barycenter position of the interference signal for the OFF data, $\bar{y}_{\mathrm{sig}}^{\mathrm{OFF}}(i)$, is presented in Figure~\ref{fig:spatial-resolution} as a function of the ``ON-OFF'' measurement number $i$ (effectively, as a function of time). 
Strong beam pointing fluctuations are clearly 
observed. The frequency spectrum of this distribution is also presented, showing a typical $1/f$ drift noise at low frequency and three harmonic peaks at 1~Hz, 2~Hz and 2.4~Hz. 
The ``ON-OFF'' subtraction (at 5~Hz) of the raw barycenter positions $\bar{y}_{\mathrm{sig}}^{\mathrm{ON}}(i) - \bar{y}_{\mathrm{sig}}^{\mathrm{OFF}}(i)$ acts as a lock-in measurement, suppressing the low-frequency noise. However, the harmonic peaks are still present and the beam pointing fluctuations are still large with a poor spatial resolution of about $1.3 \ \mu$m. Finally, the distribution of the signal $\delta y (i)$, after correcting for beam pointing fluctuations according to Eqs.~(\ref{eq:BPcorrections}), exhibits an excellent spatial resolution 
$\sigma_y(w_{\rm{RoI}}=w/2)= 40.2 \pm 0.4$~nm
in agreement with the expected CCD shot noise limit of Eq.~(\ref{eq:shot-noise}). The average value over 4000 measurements is $\langle{\delta y}\rangle = 540 \pm 636$~pm, which is compatible with the expected zero value, with sub-nanometer accuracy. 
The frequency spectrum is flat, indicating that the residual noise is purely stochastic as expected for the quantum shot noise of the CCD camera. 
This study shows that the ultimate shot noise resolution is achieved for relatively small region of interest ($w_{\rm{RoI}} \leq w/2$) when the phase noise is negligible.

\subsubsection{Phase noise pattern}

For larger region of interest, the phase noise is not negligible anymore and must be taken into account.
Using Eq.~(\ref{eq:extinction-factor}), the intensity profile at the dark output $I_{out,i}(x,y)$ for a laser pulse $i$ is 
then given by  
\begin{equation}
I_{out,i}(x,y) = I_{\rm in}(x-\bar{x}_i,y-\bar{y}_i) \times  (\delta a^2 + \delta \phi^2(x,y) ) \,,
\label{eq:sig-phase-noise}
\end{equation}
where $\bar{x}_i$ and $\bar{y}_i$ are the barycenters of the intensity profile of the incident pulse $i$, and $\delta \phi^2(x,y)$  is the matrix of the phase noise pattern. The barycenters $\bar{x}_i$ and $\bar{y}_i$ fluctuate from shot to shot with standard deviation $\sigma_b$.
It is assumed that the beamsplitter asymmetry $\delta a$ is uniform, and independent of the transverse position. 
The first term $I_S=I_{\rm in}(x-\bar{x}_i,y-\bar{y}_i) \times  \delta a^2$ in Eq.~(\ref{eq:sig-phase-noise}) corresponds to the signal intensity profile as measured in Fig.~\ref{fig:extinction-max} where $\delta a$ has been increased by rotating the beam polarization. The second term $I_B = I_{\rm in}(x-\bar{x}_i,y-\bar{y}_i) \times \delta \phi^2(x,y)$ corresponds to the phase noise pattern intensity profile as measured in Fig.~\ref{fig:rotated-polar} at maximum extinction ($\delta a \ll 1$). 
In the absence of beam pointing fluctuations, the beam profile is stable ($\sigma_b = 0$) and we can arbitrarily set $\bar{x}_i=\bar{y}_i=0$. In this case, the phase noise profile $I_B$ is equivalent to a constant offset of the intensity profile, and the spatial resolution  is only limited by the shot noise. However, in the presence of beam pointing fluctuations, the intensity profile of the incident beam, which is shifted by $\bar{x}_i, \bar{y}_i$, is now multiplied by the phase noise matrix $\delta \phi^2(x,y)$, which is independent of the beam position, and is fixed with respect to the optical elements of the interferometer. Same properties apply to the back-reflections. As a consequence, the shift of the signal intensity profile is not anymore correctly measured by the shift of the back-reflection and the spatial resolution is degraded.

This effect is well observed by measuring the spatial resolution $\sigma_y(w_{\rm{RoI}})$ as a function of the size $w_{\rm RoI}$ of the analysis window, normalized to the width $w$ (FWHM) of the beam.
As shown in Fig.~\ref{fig:fact-merite-sigy-roi}, 
for $w_{\rm RoI} < 0.5 \times w$, the phase noise intensity $I_B$ can be considered as negligible with respect to the signal intensity $I_S$ and the spatial resolution is relatively constant and close to the expected shot noise. However the resolution becomes progressively worse when $w_{\rm RoI} $ becomes larger, due to the phase noise pattern which is no longer negligible. 
This behaviour is well reproduced by the Monte Carlo simulation which now includes normal random beam pointing fluctuations and the phase noise matrix $\delta \phi^2(x,y)$ measured in Fig.~\ref{fig:fact-merite-sigy-roi}.

Therefore the beam pointing fluctuations combined to the phase noise are the  main limitation for the spatial resolution $\sigma_y$. It requires to reduce the size of the analysis window $w_{\rm RoI}$ in order to achieve the ultimate  shot noise resolution. 
However, the reduction in size of the analysis window tends to decrease the capacity to measure the displacement of the barycenter. By defining $\langle \delta y \rangle(w_{\rm RoI})$ as
\begin{equation}
\langle \delta y \rangle(w_{\rm RoI}) = \frac{\iint_{-w_{\rm RoI}/2}^{w_{\rm RoI}/2} I_{\rm out}(x,y) \times y \ dx \ dy}{\iint_{-w_{\rm RoI}/2}^{w_{\rm RoI}/2} I_{\rm out}(x,y) \ dx \ dy},
\end{equation}
we then define the efficiency $\epsilon_s(w_{\rm RoI})$ of measuring the signal, as 
\begin{equation}
\epsilon_s(w_{\rm RoI}) = \frac{\langle \delta y \rangle(w_{\rm RoI})}{\langle \delta y \rangle}.
\end{equation}
The dependence of $\epsilon_s(w_{\rm RoI})$ on the size $w_{\rm RoI}$ of the analysis window, as measured in the data, and as numerically calculated for a Gaussian transverse profile of the beam,  is shown in Fig.~\ref{fig:fact-merite-sigy-roi}. 
For $w_{\rm RoI} = 0.5 \times w$, the calculated efficiency is $\epsilon_s = 0.12$. The experimental sensitivity is maximum when the ratio
\begin{equation}
\xi(w_{\rm RoI}) = \epsilon_s(w_{\rm RoI}) \times \sigma_y(\mathrm{shot \ noise}) / \sigma_y (w_{\rm RoI}) 
\label{eq:figure-of-merit}
\end{equation}
is maximum. This ratio, equivalent to a figure of merit, is presented in Fig.~\ref{fig:fact-merite-sigy-roi} as a function of the size $w_{\rm RoI}$ of the analysis window and is maximum for $w_{\rm RoI} = 0.8 \times w$ where $\epsilon_s = 0.3$ and 
$\sigma_y(w_{\rm{RoI}}=0.8\times w) = 60$~nm.

The phase noise induced by possible beamsplitter and mirror instabilities is another possible source of systematics. 
As already mentioned, the interference pattern and the extinction factor in the dark output of a Sagnac interferometer are unmodified in the presence of a mirror or beamsplitter translation. Therefore, we are only sensitive to rotation instabilities. The direct monitoring of the extinction pattern allows us to control and correct any rotation drift at the level of tens of nanometers. 
But noise induced by smaller rotation instabilities is still present. However, as already discussed, the signal is measured by applying an ``ON-OFF'' subtraction method at 5~Hz, which acts as a lock-in measurement of a signal at the definite 5~Hz frequency. 
Therefore, only rotation instabilities at 5~Hz can contaminate the DeLLight signal, and any instabilities at other frequencies can be identified and suppressed in the frequency domain.
Finally, an important experimental test to distinguish a synchronous noise from a real DeLLight signal is to delay the pump pulse by a few picoseconds, in order to cancel the signal induced by the interaction with the pump. 

The second prototype configuration with focusing in the interferometer is much more sensitive to the turbulence of the air, which generates instabilities of the extinction pattern. This induces fluctuations of the phase noise pattern which deteriorates the spatial resolution (as previously discussed). 
Therefore this measurement must be carried out at lower pressure. This will be done in the near-future in the vacuum chamber, which is currently under development.

\subsection{Current sensitivity of the prototype}

The experimental parameters limiting the sensitivity of the experiment, given by Eq.~(\ref{eq:Nsd-sensitivity-3}), are the spatial resolution $\sigma_y$, the efficiency $\epsilon_s$, the extinction factor $\mathcal{F}$ of the interferometer, the focal length $f$ for the focusing of the probe, and the waist at focus $w_0$ of the laser beam. 
The current values of each parameter, obtained with the prototype, are summarized in Table~\ref{tab:sensitivity}. 
They are compared to the project goals, which correspond to a 1-sigma sensitivity per square root of number of days of measurement to detect the expected QED vacuum nonlinear index $n_{2,{\rm QED}}=1.6 \times 10^{-33}$~cm$^2$/W (as discussed in Section~\ref{sec:sensitivity}).
The extrapolated sensitivity of the current prototype is $n_2 = 7.4 \times 10^{-29}$~cm$^2$/W, about four orders of magnitude above the expected QED signal.
The corresponding factor to be gained is also listed separately for each parameter in the last column.

It is worth noting that, as discussed in the previous sections, the listed parameters are actually correlated and limited by two common experimental issues: the phase noise and the beam pointing fluctuations, which must both be reduced by a factor of 10 in order to reach the project goals. 
The phase noise will be reduced thanks to the current development of new beamsplitter, mirrors and lenses with superpolished surface qualities. The beam pointing fluctuations will be reduced thanks to the installation of a beam stabilization system in the LASERIX facility.
Also the shot noise, which is the ultimate limitation of the spatial resolution $\sigma_y$, will be reduced to the project goal thanks to the use of a new CCD camera with ten times the full well capacity. 
Finally it is clear that the probe waist at focus $w_0$ is a key parameter since the sensitivity scales as $\left(w_0^{2}+W_{0}^{2}\right)^{-3/2}$ ($W_{0}$ being the waist of the pump at focus). 
The diameter $w$ of the collimated probe beam in the interferometer is inversely proportional to $w_0$. Therefore, reducing $w_0$ requires to maintain the performances of the interferometer with a larger beam diameter. For instance $w \simeq 25~{\rm mm}$ for $w_0=5 \ \mathrm{\mu}$m and a focal length $f = 500$~mm for the probe beam. A possible approach to reduce by a factor of two the diameter $w$ of the probe beam, while maintaining a small waist at focus, is to double the frequency of the incident laser pulse, using second harmonic generation (from $\lambda = 800$~nm to $\lambda = 400$~nm) before entering  the interferometer.

\begin{table}[!h]
\centering
\begin{tabular}{c|c|c|c|c}
\hline
\hline
										& Prototype   			&      DeLLight  	&  Sensitivity  	  \\
										&    					&      Goal  			&  Gain 	  \\
\hline
\hline
Spatial resolution $\sigma_y$  		&  $60$~nm     		&  $10$~nm    		& 6		\\
Efficiency $\epsilon_s$  				&  $0.3$    				&  $1.0$ 		& 3			\\
Extinction factor $\mathcal{F}$ 		&  $5 \times 10^{-4}$  	&  $5 \times 10^{-6}$ & 10  				\\
Focal length $f$						&  $100$ mm			&  $500$ mm		& 5		\\
Probe waist at focus $w_0$	&  $25 \ \mu$m 		&  $5 \ \mu$m   & 	47		 \\
\hline
\hline
$n_2$ $1\sigma$-sensitivity 						& $7.4 \times 10^{-29}$ 	 & $1.8 \times 10^{-33} $ & $4 \times 10^4$  \\
$({\rm cm}^{2}/{\rm W} \times \sqrt{T_{obs}(\mathrm{days})})$   &  &  & \\
\hline
\hline
\end{tabular}
\caption{Summary of the values of the five experimental parameters which limit the sensitivity of the experiment. Values obtained with the prototype are compared to the target values of the final design.  The expected sensitivity, calculated from Eq.~(\ref{eq:Nsd-sensitivity-3}), is also calculated using results of the prototype and for the final design. The fourth column indicates the required sensitivity gain, for each parameter and in total, to reach the DeLLight specifications.
Note that, in calculating the sensitivity gain when decreasing the probe waist at focus, we assume a fixed pump waist at focus of $5\,\mu{\rm m}$.
The spatial resolution and efficiency of the prototype correspond to the values obtained with a region of interest $w_{\rm{RoI}}=0.8\times w$, which maximizes the figure of merit $\xi(w_{\rm RoI})$ defined in Eq.~(\ref{eq:figure-of-merit}).}
\label{tab:sensitivity}
\end{table}

\section{Conclusion}

Quantum electrodynamics predicts that the speed of light in vacuum must be reduced when the vacuum is stressed by intense electromagnetic fields. This has not yet 
been observed 
and remains one of the most intriguing experimental predictions of QED at macroscopic scales. 
To date, the most sensitive tests involve the search for a nonlinear vacuum birefringence induced by an external magnetic field. 

In this article, we have presented 
a new experimental method which directly exploits the change in the vacuum refractive index, rather than the associated birefringence.
The idea is to measure the refraction of a focused probe laser pulse crossing the transverse vacuum index gradient induced by a focused intense pump pulse. 
The associated nonlinear index is $n_{2,{\rm QED}}=1.6 \times 10^{-33}$~cm$^2$/W, as predicted by the Euler-Heisenberg model of nonlinear electrodynamics derived from QED. 
The DeLLight experiment has recently been installed in the LASERIX facility at IJCLab, designed to deliver pulses with energy of 2.5~J and durations 30~fs, at a repetition rate of 10~Hz. 
We have shown that, if pump and probe are both focused with a minimum waist of $5 \mu$m (corresponding to a maximum intensity of $\sim 3 \times 10^{20}$~W/cm$^2$), the expected refraction angle is 0.13~prad. 
The refraction of the probe is measured via a Sagnac interferometer, the signal corresponding to a transverse displacement of the intensity profile in the dark output of the interferometer. 
We have shown that the interferometric method amplifies the signal by a factor which scales as the inverse of the square root of the extinction factor $\mathcal{F}$. 
If $\mathcal{F}=0.5\times10^{-5}$, the measured displacement 
is amplified by a factor of 250 and is calculated to be about 15~pm when using a focal length of 500~mm for the focusing of the probe. 
The signal of displacement is measured by alternating laser shots with and without interactions between the pump and the probe pulse (ON and OFF measurements), and by calculating the ON-OFF signal. 
We have shown that this signal can be detected with LASERIX at a 5-sigma confidence level with about one month of collected data if we maintain a spatial resolution of $\sigma_y = 10$~nm for measuring the barycenter of the intensity profile. This corresponds to the typical lowest shot noise of current available CCD camera. 

The DeLLight experiment will be run in a first phase, using the LASERIX facility operating at around 80~TW. For a second phase, the experiment could be run using more intense PW or multi-PW lasers coming to operation. As shown in Eq.~(\ref{eq:Nsd-sensitivity-3}), the sensitivity of the experiment is proportional to the energy of the laser pulses and, for a given duration of measurement, proportional to the square root of the repetition rate of the laser shots. Thus a very suitable laser is the L3 laser system named HAPLS (High-Repetition-Rate Advanced Petawatt Laser System)~\cite{HAPLS} installed at the ELI-Beamlines Center (Czech Republic), which is designed to deliver pulses with energy of at least 30~J and duration 30~fs (1~PW) at a repetition rate of 10~Hz. Using the HAPLS laser, the DeLLight displacement signal is expected to be about 0.2~nm, 12 times larger than the expected signal with LASERIX, and could be detected at 5-sigma in only 6 hours of collected data. Other lasers coming to operation, as ELI-NP facility (Romania)~\cite{ELI-NP} or SULF facility (China)~\cite{SULF}, will deliver 10~PW pulses at 1 shot per minute (0.017~mHz). With this intensity, the DeLLight signal is expected to be on the order of 1~nm. However, because of their very small repetition rate, the achievable sensitivity is slightly lower than that achievable with HAPLS, and the drift noise (which is inversely proportional to the repetition rate) becomes almost three order of magnitude larger. 

A first Sagnac interferometer has been developed in the LASERIX facility and we have measured the critical parameters (in addition to the laser intensity), which are the extinction of the interferometer, the spatial resolution, and the transverse size of the focused pulses in the interferometer.
The achieved sensitivity of this first prototype is about four orders of magnitude above the sensitivity goal. However, these results have been obtained with a precursory setup, and it is shown that the final DeLLight specifications can be achieved by improving the surface quality of the optics, by improving the stability of the beam and its focus quality, and by using CCD cameras with the best available charge storage capacity per unit surface.

\section*{Acknowledgments}
This work is funded by the French National Research Agency via Grant No. ANR-18-CE31-0005-01.


\begin{appendices}


\section{Accounting for a non-zero tilt angle}
\label{Appendix-A}

In this appendix we generalize the calculation presented in Sec.~\ref{sub:deflection} to one in which the tilt angle $\theta_{\rm tilt}$ between the trajectories of pump and probe is non-zero.

Let us symmetrize the descriptions of pump and probe by using coordinates in which each of them has a trajectory at angle $\theta_{\rm tilt}/2$ with respect to the $z$-axis (see Fig.~\ref{fig:pulses_angle}).  Consider first the intensity profile of the probe.  With respect to the profile in the ``old'' coordinates:
\begin{equation}
I_{\rm probe}(x,y,z) = I_{\rm in} \, {\rm exp}\left( -2\frac{x^{2}}{w_{0}^{2}}  -2\frac{y^{2}}{w_{0}^{2}}  -2\frac{z^{2}}{w_{z}^{2}} \right) \,,
\end{equation}
we effectively make a rotation of the form $x \to x\,{\rm cos}\left(\theta_{\rm tilt}/2\right) + z\,{\rm sin}\left(\theta_{\rm tilt}/2\right)$ and $z \to z\,{\rm cos}\left(\theta_{\rm tilt}/2\right) - x\,{\rm sin}\left(\theta_{\rm tilt}/2\right)$, so that we have instead:
\begin{widetext}
\begin{equation}
I_{\rm probe}(x,y,z) = I_{\rm in} \, {\rm exp}\left( -2\frac{\left(x\,{\rm cos}\left(\theta_{\rm tilt}/2\right) + z\,{\rm sin}\left(\theta_{\rm tilt}/2\right)\right)^{2}}{w_{0}^{2}} \right) {\rm exp}\left( -2\frac{y^{2}}{w_{0}^{2}} \right) {\rm exp}\left( -2\frac{\left(z\,{\rm cos}\left(\theta_{\rm tilt}/2\right) - x\,{\rm sin}\left(\theta_{\rm tilt}/2\right)\right)^{2}}{w_{z}^{2}} \right) \,.
\label{eq:I_probe_rotated}
\end{equation}
The transformation of the pump profile behaves analogously, the only differences being:
\begin{itemize}
\item the occurrence of the impact parameter $b$, so that the exponential in $y$ becomes ${\rm exp}\left(-2\left(y-b\right)^{2}/W_{0}^{2}\right)$; and
\item the rotation by $\theta_{\rm tilt}/2$ occurring in the opposite direction, so that, in effect, the terms in ${\rm sin}\left(\theta_{\rm tilt}/2\right)$ change sign.
\end{itemize}
By completing the square in $z$, Eq.~(\ref{eq:I_probe_rotated}) can be rewritten in the form,
\begin{multline}
I_{\rm probe}(x,y,z) = I_{\rm in} \, {\rm exp}\left(-2\frac{w_{0}^{2}{\rm cos}^{2}\frac{\theta_{\rm tilt}}{2} + w_{z}^{2}{\rm sin}^{2}\frac{\theta_{\rm tilt}}{2}}{w_{0}^{2}w_{z}^{2}} \left(z + \frac{{\rm cos}\frac{\theta_{\rm tilt}}{2} {\rm sin}\frac{\theta_{\rm tilt}}{2} \left(w_{0}^{2}-w_{z}^{2}\right)}{w_{0}^{2}{\rm cos}^{2}\frac{\theta_{\rm tilt}}{2} + w_{z}^{2}{\rm sin}^{2}\frac{\theta_{\rm tilt}}{2}} x\right)^{2} \right) \\
\times {\rm exp}\left(-2\frac{y^{2}}{w_{0}^{2}} \right) {\rm exp}\left(-2\frac{x^{2}}{w_{0}^{2}{\rm cos}^{2}\frac{\theta_{\rm tilt}}{2} + w_{z}^{2}{\rm sin}^{2}\frac{\theta_{\rm tilt}}{2}}\right) \,.
\end{multline}
and similarly for $\delta n(x,y,z)$.
\end{widetext}

So far we have considered only the spatial profiles of the two pulses.  Let us now consider how these vary in time.  We assume, as before, that the interaction region is short enough that there is no significant divergence of the pulses, and that they vary in time only on account of their propagation in space at a constant velocity.  In the lab frame, these velocities have magnitude $c$ and are aligned with the pulse axes; more precisely, they are, in the $(x,y,z)$-coordinate system,
\begin{eqnarray}
{\bf u}_{\rm pump} &=& c \left(-{\rm sin}\frac{\theta_{\rm tilt}}{2},0,{\rm cos}\frac{\theta_{\rm tilt}}{2}\right) \,, \nonumber \\ 
{\bf u}_{\rm probe} &=& c \left(-{\rm sin}\frac{\theta_{\rm tilt}}{2},0,-{\rm cos}\frac{\theta_{\rm tilt}}{2}\right) \,.
\end{eqnarray}
It is, however, more convenient to use a coordinate system that is attached to the probe, so that each position ${\bf r}^{\prime} = {\bf r} - {\bf u}_{\rm probe} t$ corresponds to a fixed point within the probe pulse.  Then only the pump pulse and its associated index variation $\delta n$ are actually moving, and its velocity in this coordinate system will be
\begin{equation}
{\bf u}_{\rm pump}^{\prime} = {\bf u}_{\rm pump} - {\bf u}_{\rm probe} = 2 c \, {\rm cos}\frac{\theta_{\rm tilt}}{2} \, \hat{\bf z} \,.
\end{equation}
In this coordinate system, the infinitesimal distance travelled by a ray within the probe becomes simply $c\,{\rm d}t$, and so Fermat's principle of Eq.~(\ref{eq:deflection_rate}) becomes
\begin{equation}
\frac{{\rm d} \theta_{y}}{{\rm d} t} = c \, \partial_{y} \left[\delta n\left({\bf r}^{\prime} - {\bf u}^{\prime} t\right) \right]  \,,
\end{equation}
which upon integration gives
\begin{eqnarray}
\delta\theta_{y}\left({\bf r}^{\prime}\right) &=& c\, \partial_{y} \left[ \int_{-\infty}^{+\infty} \delta n \left({\bf r}^{\prime} - {\bf u}^{\prime} t \right) \, dt \right] \nonumber \\
&=& \frac{c}{\left|{\bf u}^{\prime}\right|} \, \partial_{y} \left[ \int_{-\infty}^{+\infty} \delta n \left(x, y, z^{\prime} - Z \right) \, dZ \right] \,,
\label{eq:delta-theta_Z-integral}
\end{eqnarray}
where in the last equality we have used the integration variable $Z = \left|{\bf u}^{\prime}\right| t$.  Note that the factor of $1/\left|{\bf u}^{\prime}\right|$ accounts for the interaction time between the two pulses, generating the $1/{\rm cos}\left(\theta_{\rm tilt}/2\right)$ factor that leads to a third power in the numerator of $r_{\rm tilt}$ (see Eq.~(\ref{eq:rtilt})) instead of the fourth power characterizing the nonlinear index (see Eq.~(\ref{eq:n2})).  

From Eq.~(\ref{eq:delta-theta_Z-integral}) we see that, as before, we can integrate directly over $Z$ and that $\delta\theta_{y}$ is a function only of $x$ and $y$.  Performing this integration, we have
\begin{widetext}
\begin{equation}
\delta\theta_{y}(x,y) = \partial_{y}\left[ \frac{\delta n_{\rm max}}{2\,{\rm cos}\frac{\theta_{\rm tilt}}{2}} \sqrt{\frac{\pi}{2}} \frac{W_{0}W_{z}}{\sqrt{W_{0}^{2}\,{\rm cos}^{2}\frac{\theta_{\rm tilt}}{2} + W_{z}^{2}\,{\rm sin}^{2}\frac{\theta_{\rm tilt}}{2}}} \, {\rm exp}\left(-2\frac{y^{2}}{W_{0}^{2}} \right) {\rm exp}\left(-2\frac{x^{2}}{W_{0}^{2}{\rm cos}^{2}\frac{\theta_{\rm tilt}}{2} + W_{z}^{2}{\rm sin}^{2}\frac{\theta_{\rm tilt}}{2}}\right) \right] \,.
\end{equation}
We now wish simply to write $\delta n_{\rm max}$ in terms of the energy and the dimensions of the pump pulse.  Using Eq.~(\ref{eq:n2}) and the Gaussian profile of the pulse, we have
\begin{equation}
\delta n_{\rm max} = n_{2,{\rm max}} \, c \, {\rm cos}^{4}\frac{\theta_{\rm tilt}}{2} \, \left(\frac{2}{\pi}\right)^{3/2} \frac{\mathcal{E}}{W_{0}^{2}W_{z}} \,,
\end{equation}
and so
\begin{multline}
\delta\theta_{y}(x,y) = \partial_{y}\left[ \frac{1}{\pi} \, c \,n_{2,{\rm max}} \, {\rm cos}^{3}\frac{\theta_{\rm tilt}}{2} \, \frac{\mathcal{E}}{W_{0}} \, \frac{1}{\sqrt{W_{0}^{2}\,{\rm cos}^{2}\frac{\theta_{\rm tilt}}{2} + W_{z}^{2}\,{\rm sin}^{2}\frac{\theta_{\rm tilt}}{2}}} \right. \\
\left. \times \, {\rm exp}\left(-2\frac{\left(y-b\right)^{2}}{W_{0}^{2}} \right) {\rm exp}\left(-2\frac{x^{2}}{W_{0}^{2}{\rm cos}^{2}\frac{\theta_{\rm tilt}}{2} + W_{z}^{2}{\rm sin}^{2}\frac{\theta_{\rm tilt}}{2}}\right) \right] \,.
\end{multline}
\end{widetext}

The last step is to average $\delta\theta_{y}(x,y)$ over all $x$ and $y$, using $I_{\rm probe}$ as a weighting function.  Once again, the integral over $z$ in the numerator is cancelled by the same integral in the denominator, so we can ignore the exponential factor in $z$ and consider only those in $x$ and $y$, i.e., we calculate
\begin{equation}
\left\langle \delta\theta_{y} \right\rangle = \frac{\int\int {\rm d}x \, {\rm d}y \, \delta\theta_{y}(x,y) \, I_{\rm probe}^{\prime}(x,y)}{\int\int {\rm d}x \, {\rm d}y \, I_{\rm probe}^{\prime}(x,y)} \,,
\end{equation}
where we have defined
\begin{eqnarray}
I_{\rm probe}^{\prime}(x,y) &=& I_{\rm in} \, {\rm exp}\left(-2\frac{y^{2}}{w_{0}^{2}}\right) \times \nonumber \\ 
&& \!\!\!\! {\rm exp}\left(-2\frac{x^{2}}{w_{0}^{2}{\rm cos}^{2}\frac{\theta_{\rm tilt}}{2} + w_{z}^{2}{\rm sin}^{2}\frac{\theta_{\rm tilt}}{2}}\right) \,.
\end{eqnarray}
The $y$-integral is decoupled from the rotation in the $xz$-plane, and therefore works out exactly as before.  Only the $x$-integral is altered by $\theta_{\rm tilt}$.  The result is
\begin{equation}
\left\langle \delta\theta_{y} \right\rangle = \frac{c \, n_{2,{\rm max}}}{4 \pi \sqrt{e}} \, \frac{\mathcal{E}}{b_{\rm opt}^{3}} \, r_{\rm tilt} \, \frac{b}{b_{\rm opt}} \, {\rm exp}\left\{ \frac{1}{2} \left( 1 - \left(\frac{b}{b_{\rm opt}}\right)^{2} \right) \right\} \,,
\end{equation}
where we have defined
\begin{equation}
b_{\rm opt} = \frac{1}{2} \sqrt{ W_{0}^{2} + w_{0}^{2} }
\end{equation}
and
\begin{equation}
r_{\rm tilt} = \frac{{\rm cos}^{3}\left(\frac{\theta_{\rm tilt}}{2}\right)}{\sqrt{1 + \left(R^{2}-1\right) \, {\rm sin}^{2}\left(\frac{\theta_{\rm tilt}}{2}\right)}} \,, \quad R^{2} = \frac{W_{z}^{2}+w_{z}^{2}}{W_{0}^{2}+w_{0}^{2}} \,.
\end{equation}

\end{appendices}



\newpage
\begin{widetext}

\begin{figure}
\centering
  \includegraphics[width=0.48\columnwidth]{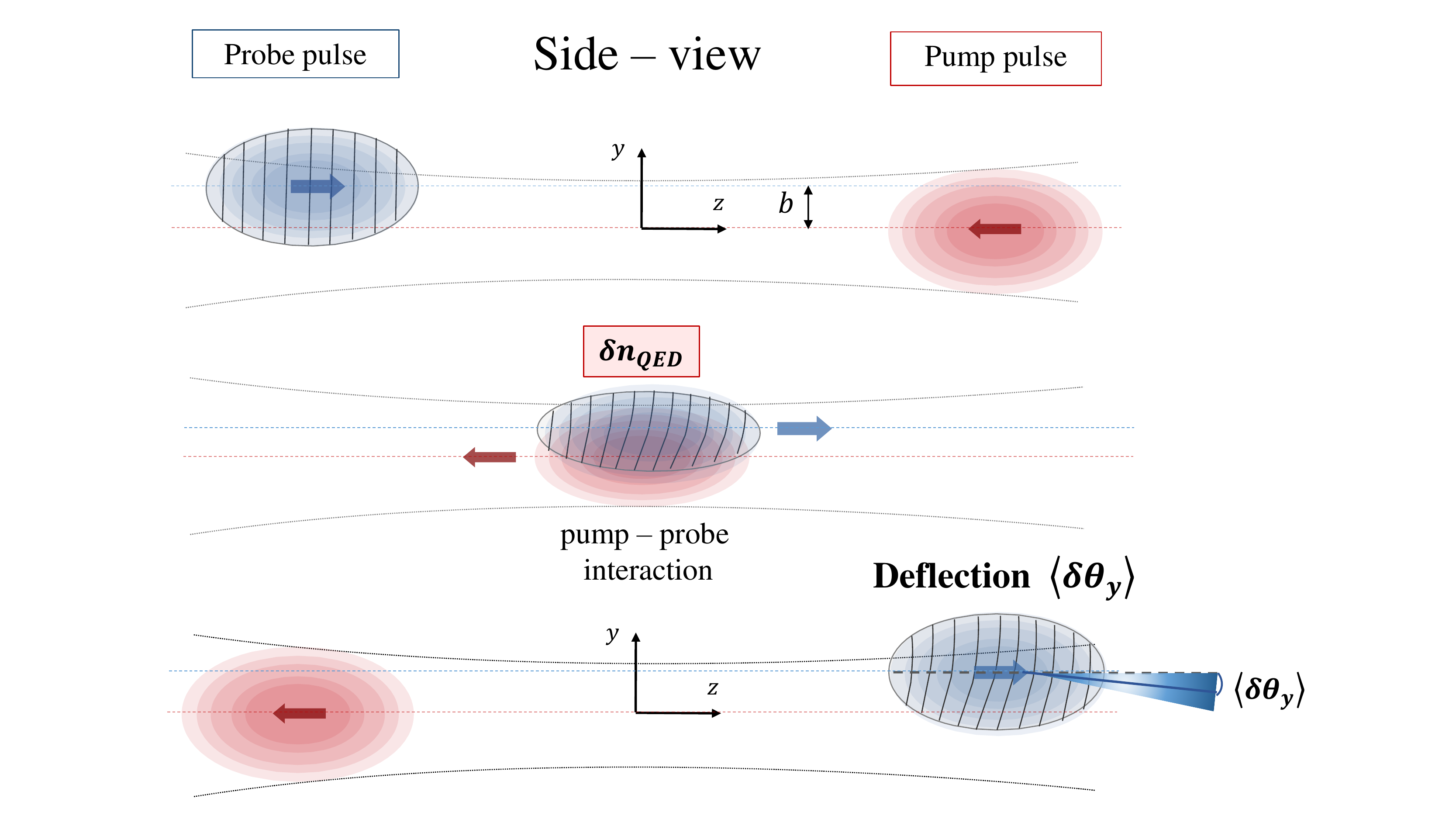} \, \includegraphics[width=0.48\columnwidth]{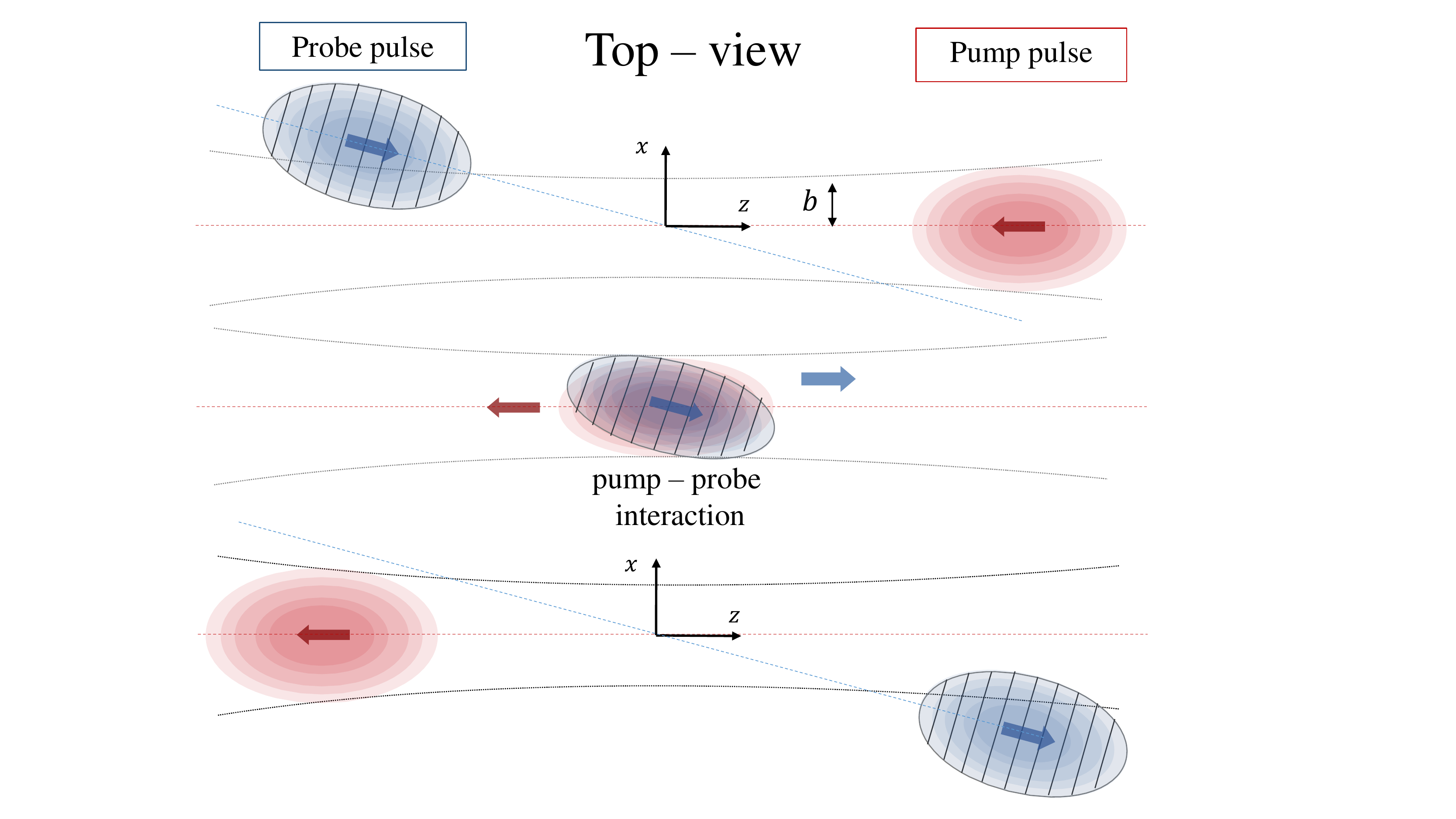}
  \caption{Schematic view of the interaction between the probe pulse (in blue) and the counter-propagating pump pulse (in red). 
The lines inside the probe pulse correspond to the wavefronts, which are gradually rotated by the vacuum index gradient induced by the pump.
(Left) Side-view: the axis of the pump beam is vertically ($y$-axis) shifted with respect to the axis of the probe beam, thus engendering an impact parameter $b$ so that the perturbation of the probe is asymmetric and the mean deflection is non-zero along the vertical $y$-axis. (Right) Top-view: in the horizontal plane (corresponding to the Sagnac interferometer $x-z$ plane), the axis of the pump beam is tilted by an angle $\theta_{tilt}$ with respect to the axis of the probe beam. The perturbation of the probe is symmetric and the mean deflection is zero along the horizontal $x$ axis.}
  \label{fig:pump-probe-refraction}
\end{figure}

\begin{figure}
\centering
  \includegraphics[scale=0.45]{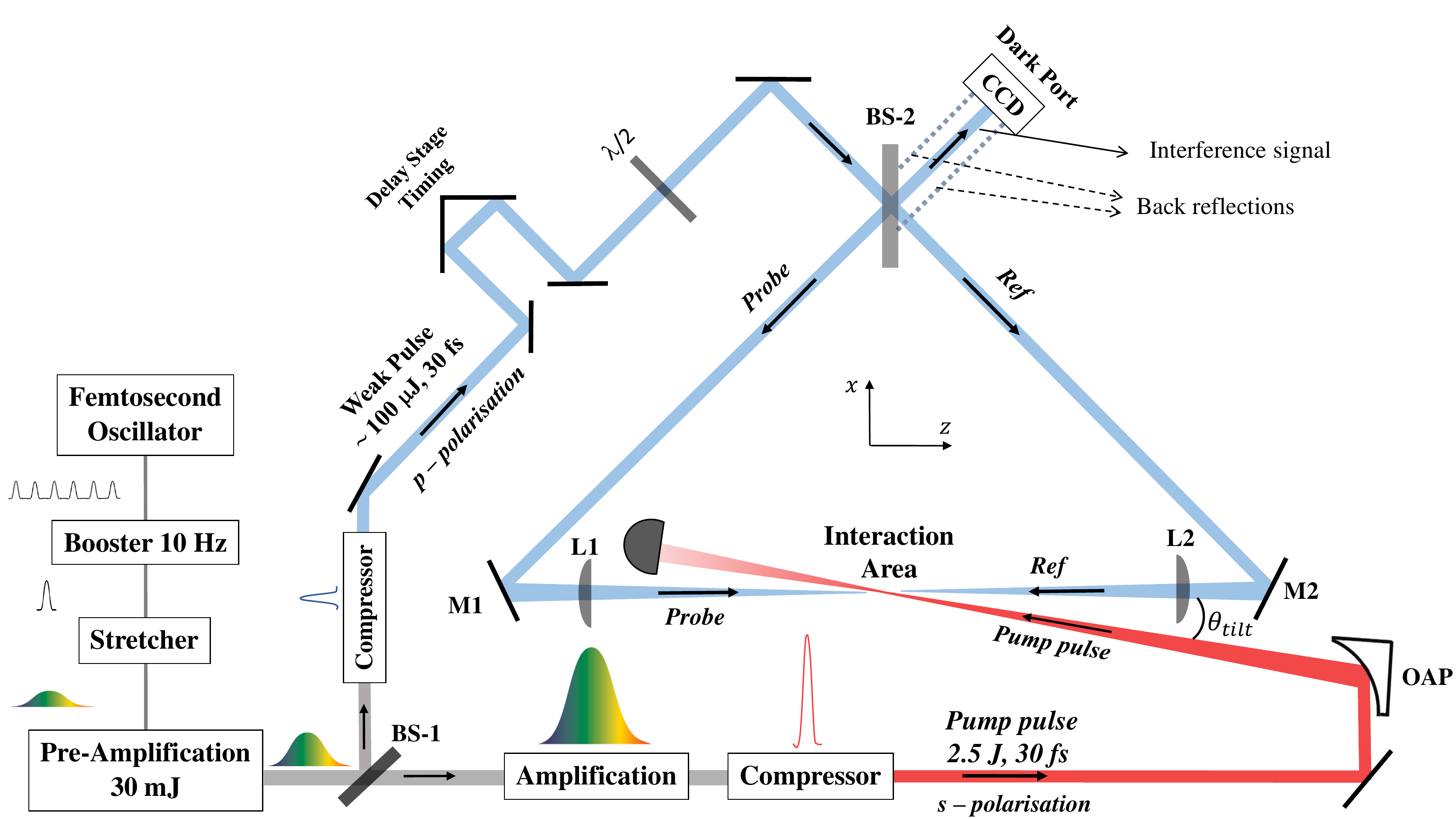}
  \caption{Schematic view of the DeLLight experimental setup (see text for details).}
  \label{fig:setup}
\end{figure}

\begin{figure}
\centering
\includegraphics[width=0.48\columnwidth]{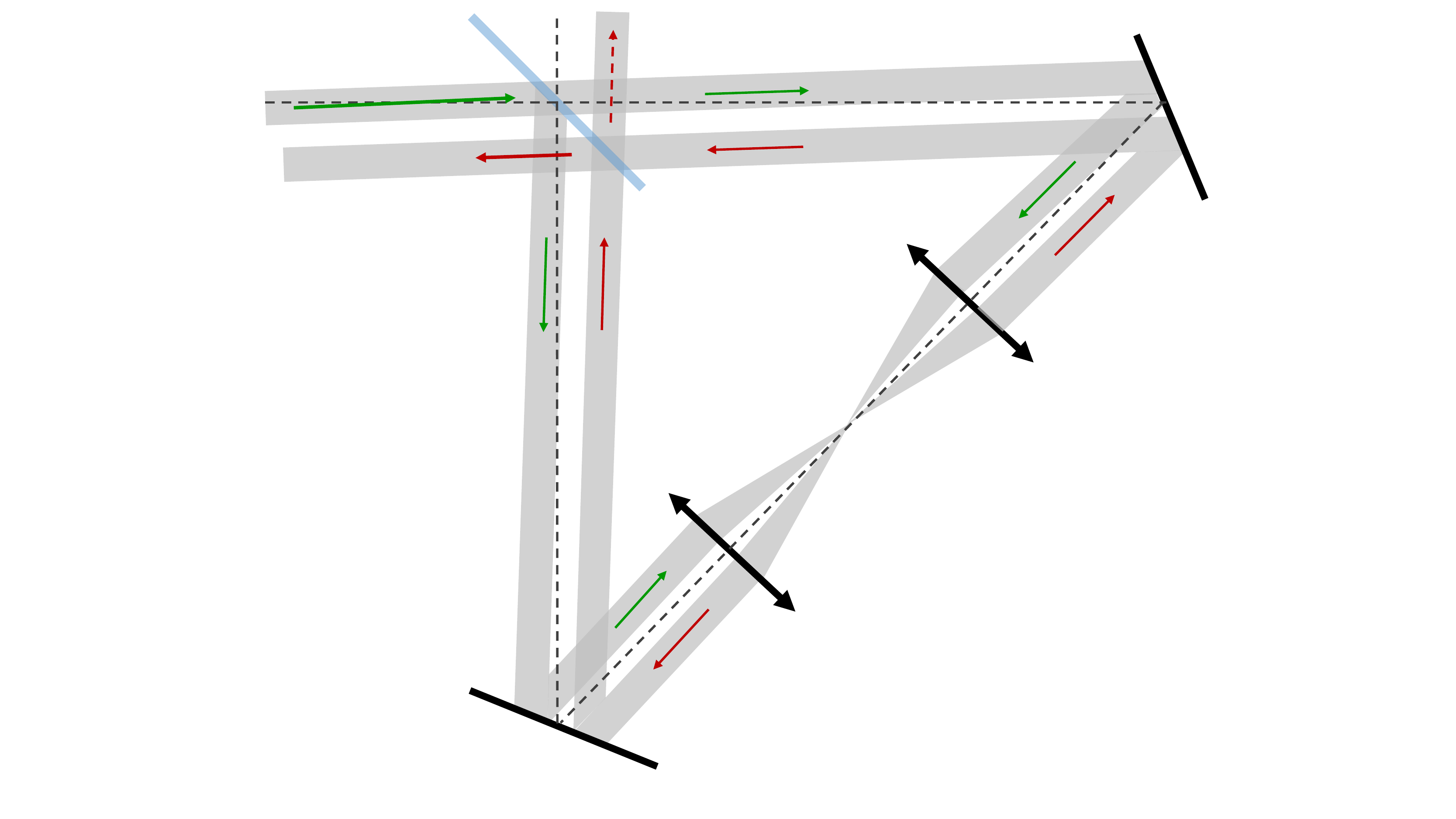} \, \includegraphics[width=0.48\columnwidth]{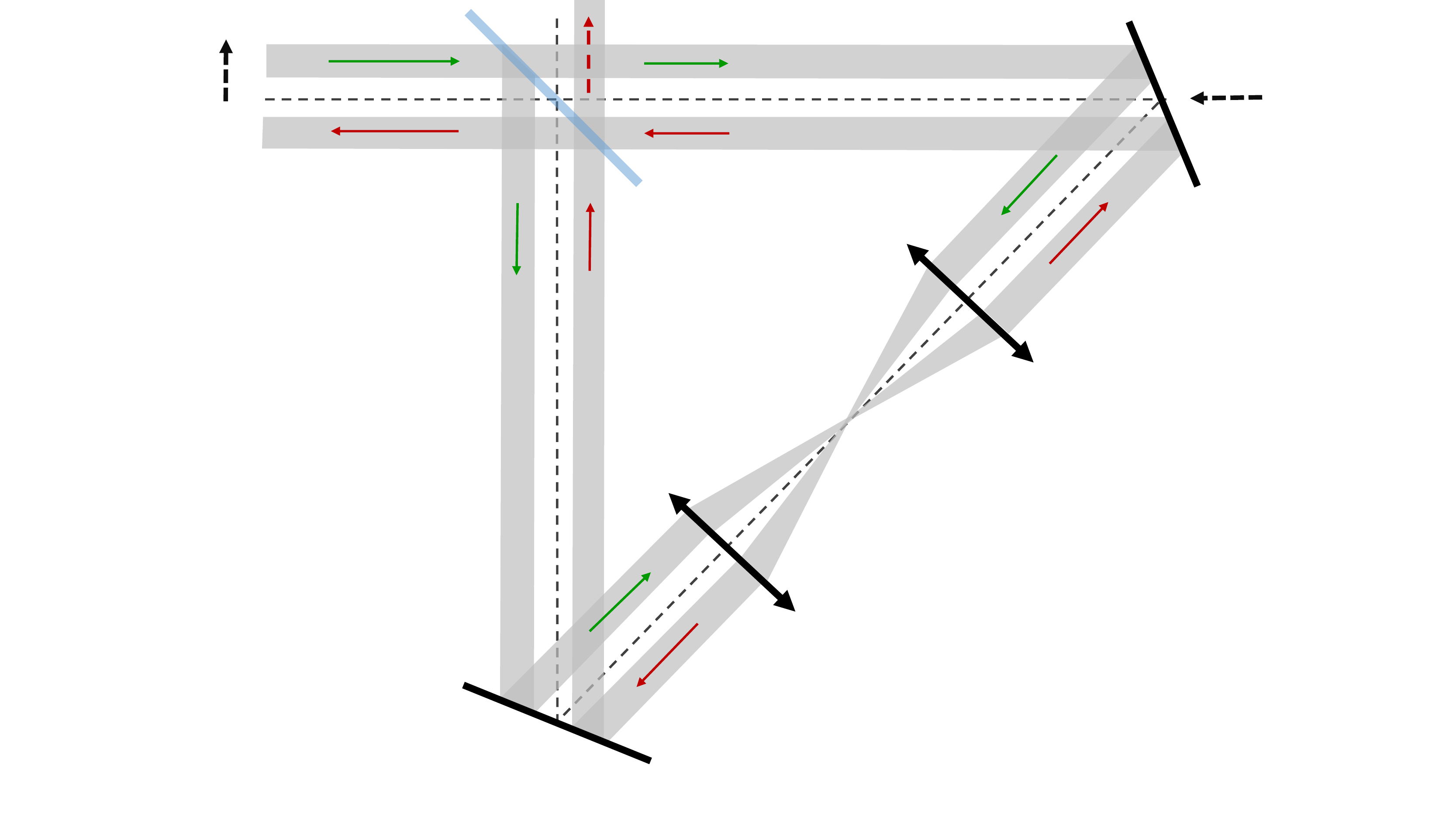}
  \caption{The extinction in the dark output of the Sagnac interferometer is insensitive to the beam pointing fluctuations. (Left) The incident beam is tilted by a small angle. (Right) The beam, or equivalently a mirror, is transversally shifted in the horizontal plane. In both cases, the extinction in the dark output of the interferometer is maintained. Only the position of the intensity profile is transversally shifted. This shift is measured and suppressed by monitoring online the same shift of the back-reflections from the rear side of the beamsplitter.}
  \label{fig:schema-beampointingfluctuation}
\end{figure}

\begin{figure}
\includegraphics[width=0.45\columnwidth]{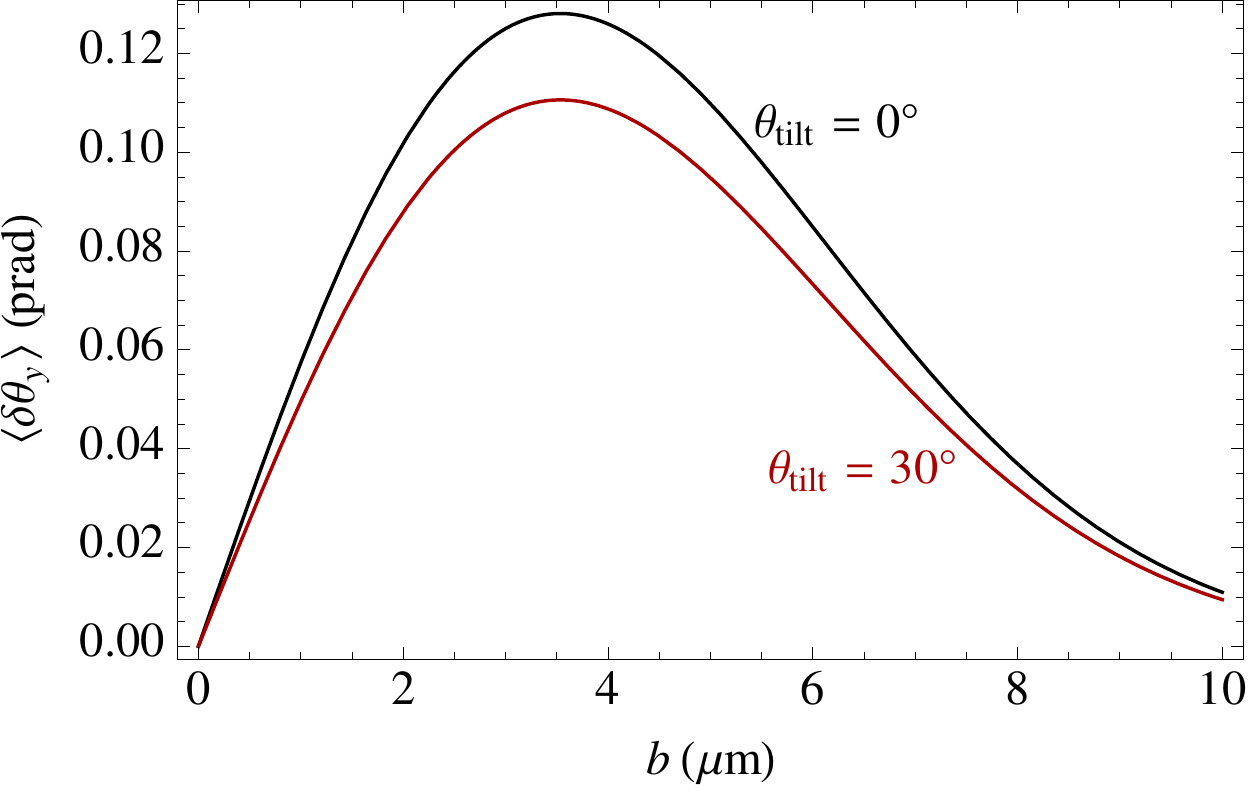} \, \includegraphics[width=0.45\columnwidth]{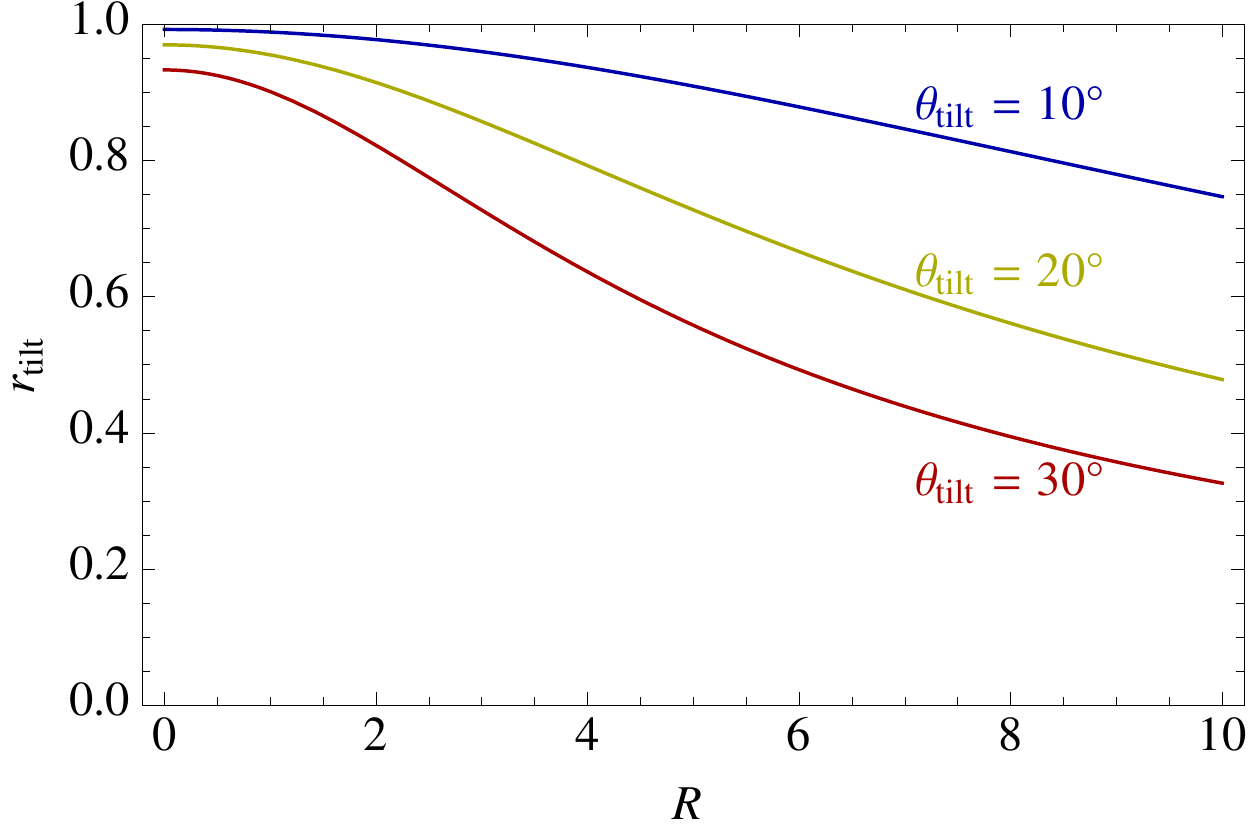}
\caption{On the left is shown the mean angular deflection of the barycenter, $\langle \delta\theta_{y} \rangle$, as a function of the impact parameter $b$ giving the shift between the focus axes of pump and probe.  The pulses are assumed orthogonally polarized, and each has a minimum waist of $5\,\mu{\rm m}$ and a duration $\Delta t_{\rm FWHM} = 30\,{\rm fs}$, while the total energy of the pump pulse is $2.5\,{\rm J}$.  The black curve corresponds to a ``head-on'' collision $\theta_{\rm tilt}=0$, while the red curve corresponds to $\theta_{\rm tilt} = 30\degree$.  On the right is shown the correction factor $r_{\rm tilt}$, for three different tilt angles, as a function of the dimensionless ratio $R$ which characterizes the elongation of the pulses (see Eq.~\ref{eq:rtilt}).
\label{fig:delta-theta-y}}
\end{figure}

\begin{figure}
\includegraphics[width=0.45\columnwidth]{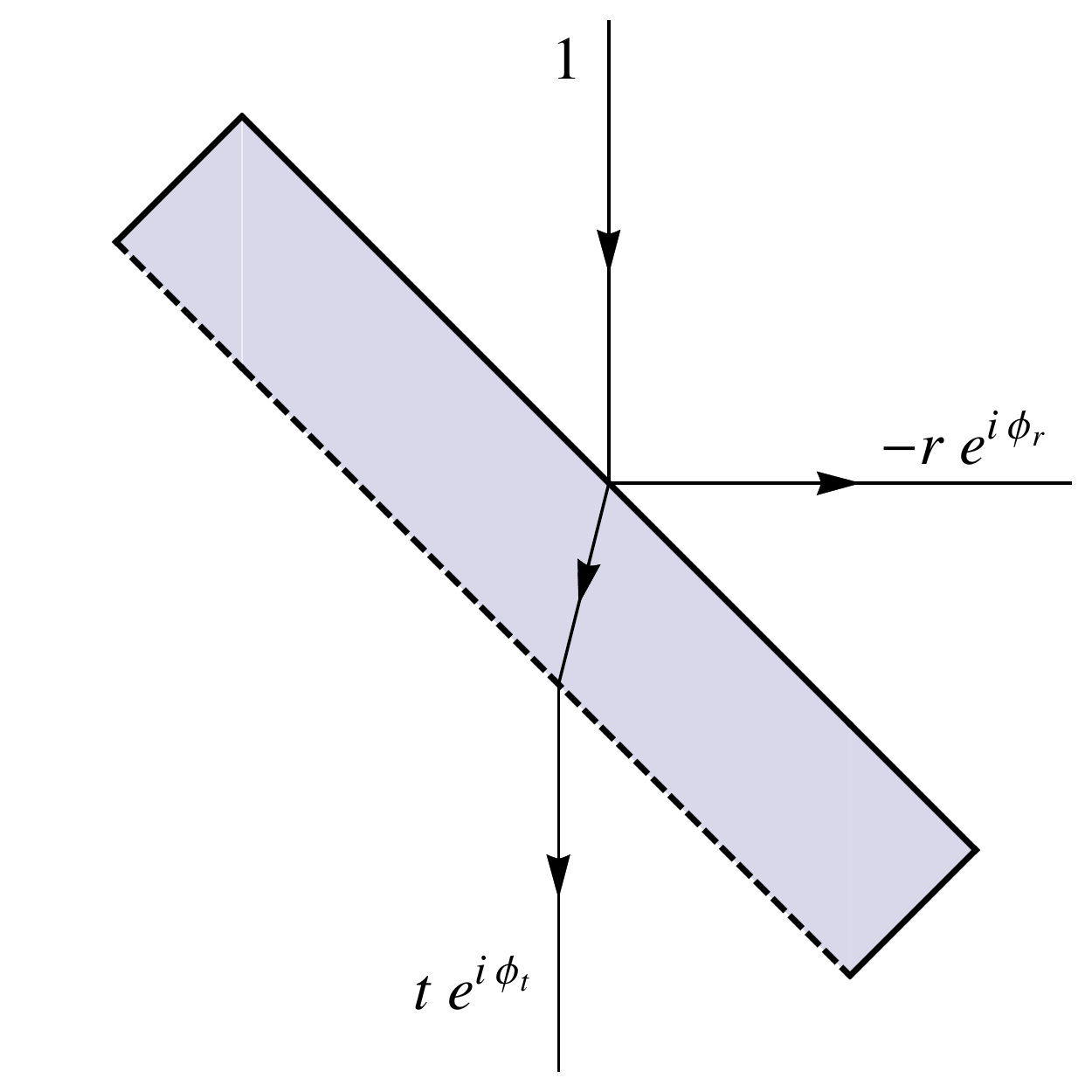} \qquad \includegraphics[width=0.45\columnwidth]{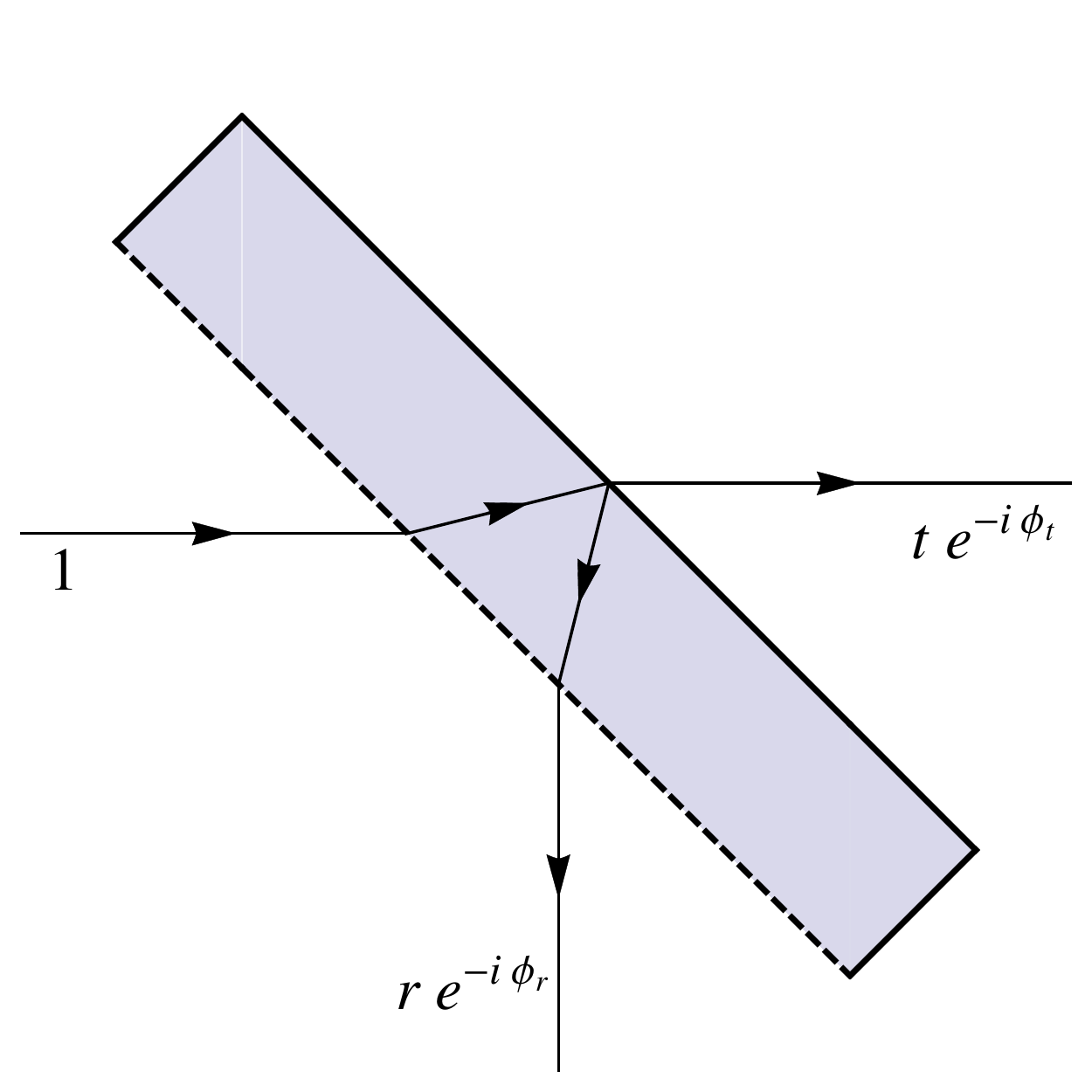}
\caption{Scattering at a lossless beamsplitter. The amplitudes of the ingoing and outgoing waves are related by the scattering matrix $S$ (see Eq.~(\ref{eq:S-matrix})).  Neglecting the overall phase $\Phi_{0}$ accumulated by all outgoing waves, as well as the reflection of rays from the anti-reflective coating (which is shown here in dashed line), the amplitudes of the outgoing waves must have the form shown here.  The two panels correspond to the scattering of rays incident on the beamsplitter from opposite sides.  These two scattering processes are characterized by the same amplitudes and phases, with some relative signs.
\label{fig:BS}}
\end{figure}
\begin{figure}
\includegraphics[width=0.45\columnwidth]{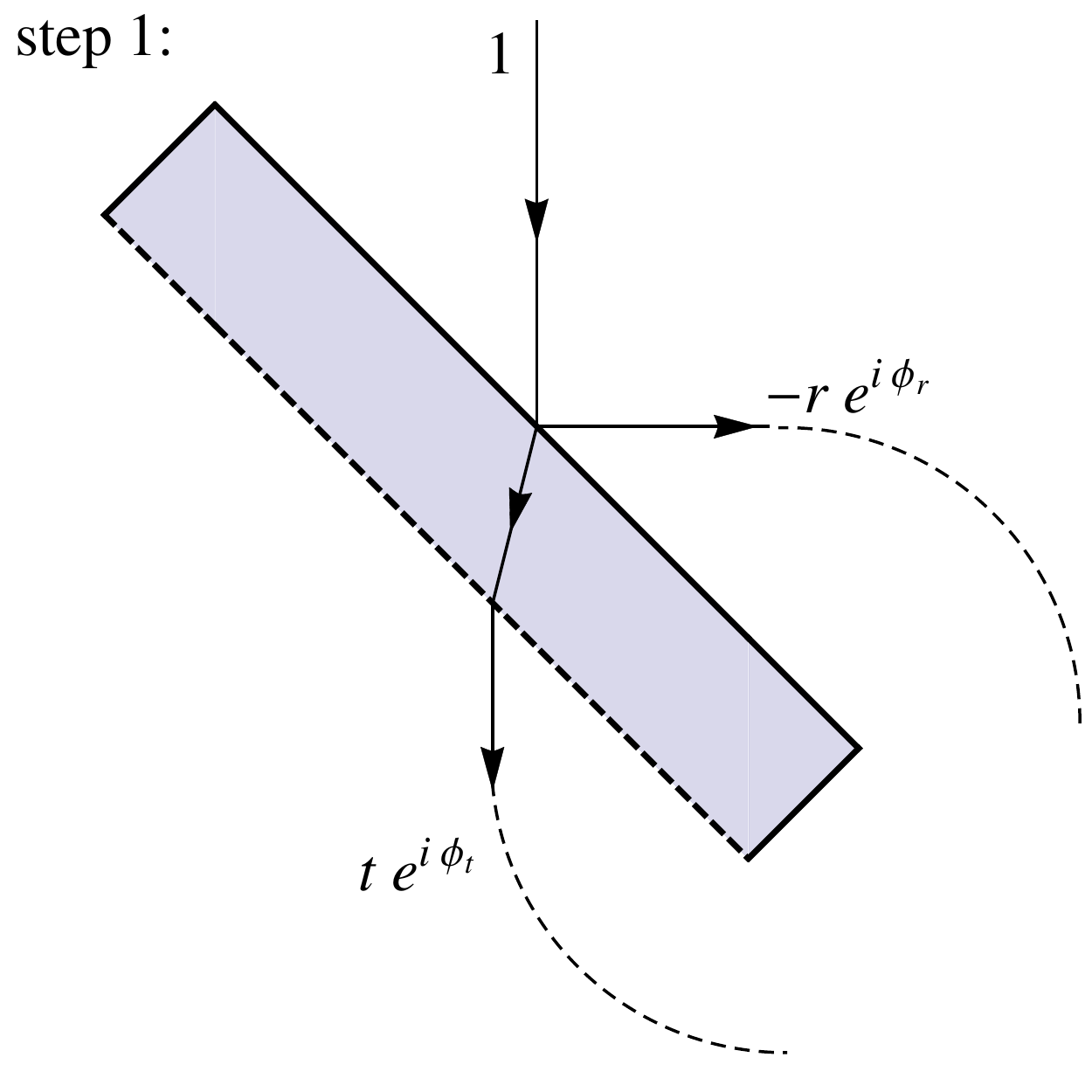} \, \includegraphics[width=0.45\columnwidth]{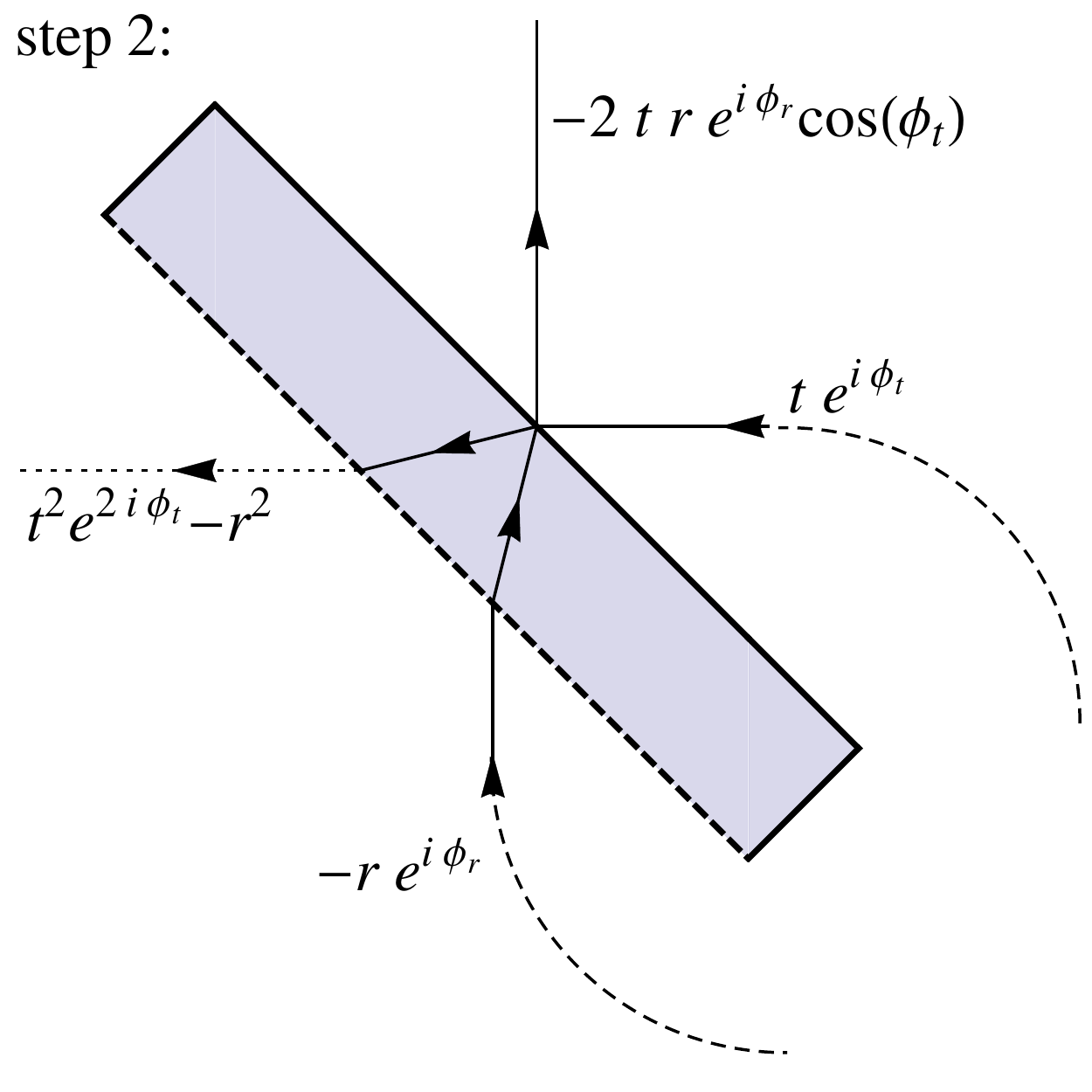}
\caption{Scattering at the beamsplitter in a Sagnac interferometer.  In the first stage, a single incident wave is split into transmitted and reflected parts, exactly as in the left panel of Fig.~\ref{fig:BS}.  In the second stage, these two parts are recombined at the same beamsplitter, this being described by a linear combination of the two panels of Fig.~\ref{fig:BS} (and recalling that the beamsplitter is symmetric under rotations in its plane, so that the same scattering amplitudes apply even though the directions of the waves have been reversed).  Assuming the initial wave has amplitude $1$, the amplitudes of the various waves are shown.  (Again, the overall phase $\Phi_{0}$ accumulated at each interaction with the beamsplitter has been suppressed for simplicity, and reflections at the anti-reflective coating have been neglected.)
\label{fig:Sagnac_BS}}
\end{figure}

\begin{figure}
\centering
  \includegraphics[scale=0.45]{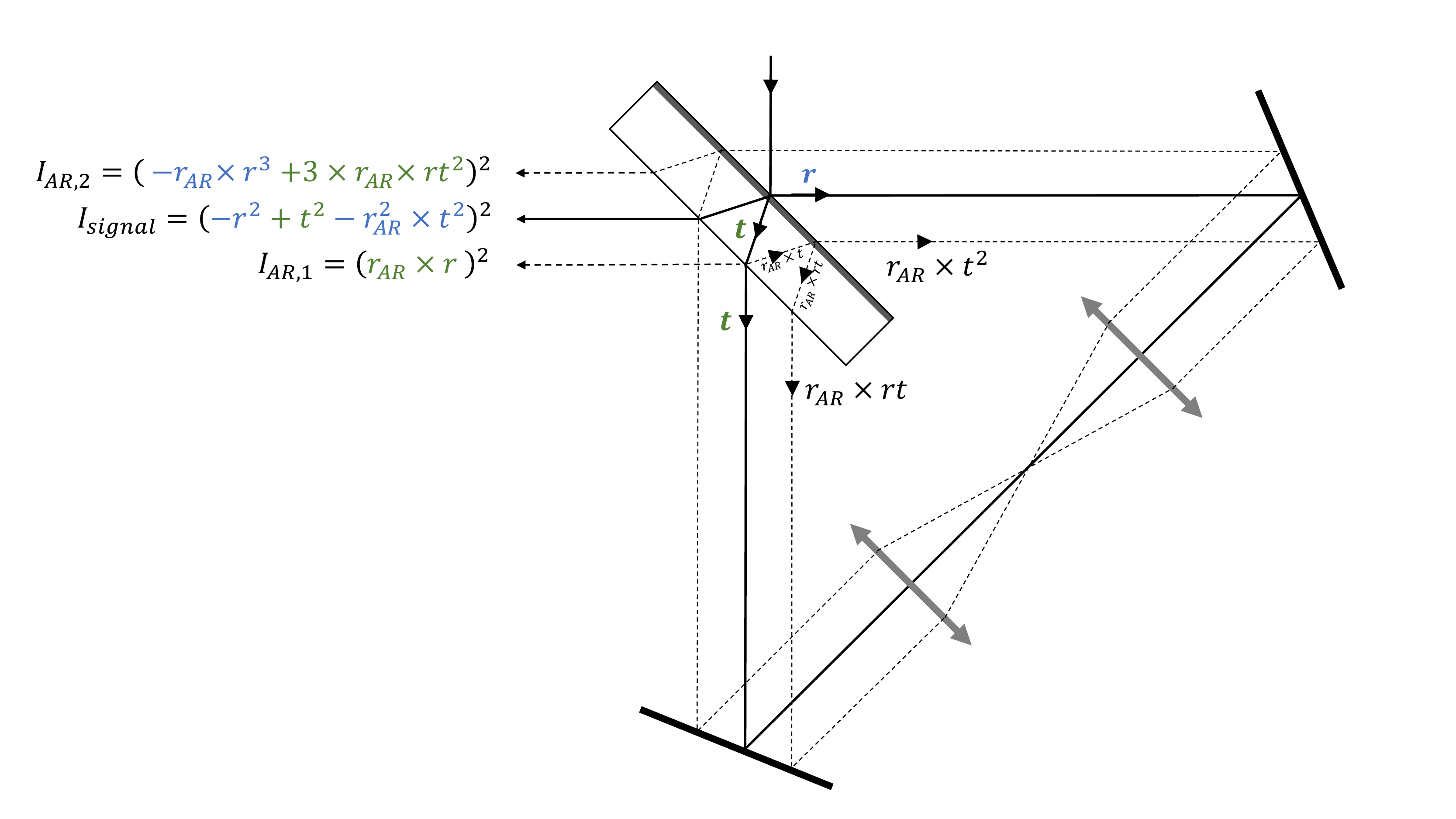}
  \caption{Ray tracing of the main back reflections on the rear side of the beamsplitter BS-2 (See Figure~\ref{fig:setup}). In blue, an odd number of reflections (from a low to a high index) corresponding to an extra phase $\pi$; in green, an even number of reflections equivalent to a null phase.
 }
  \label{fig:back-reflexions}
\end{figure}

\begin{figure}
\centering
  \includegraphics[width=0.48\columnwidth]{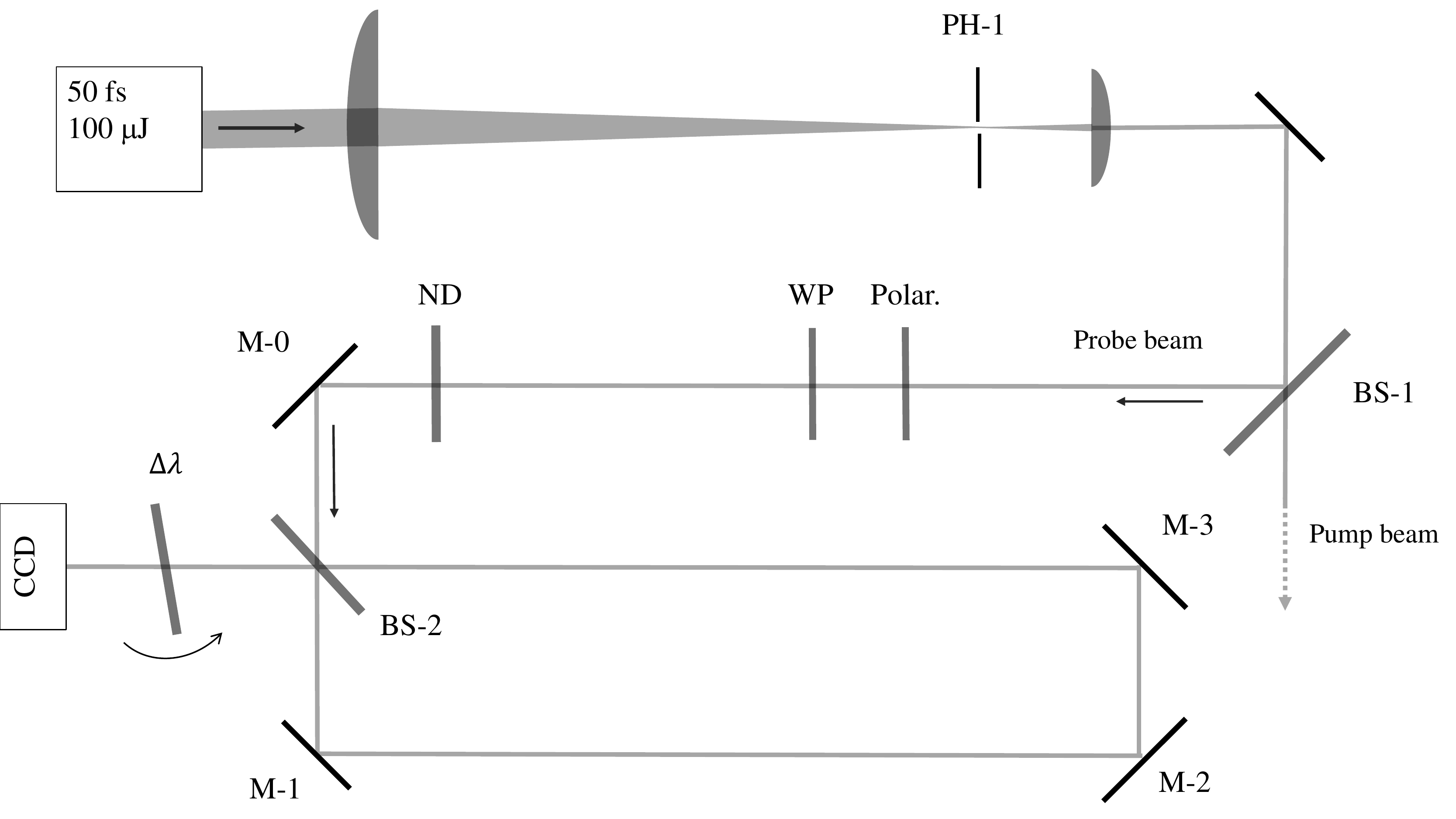} \, \includegraphics[width=0.5\columnwidth]{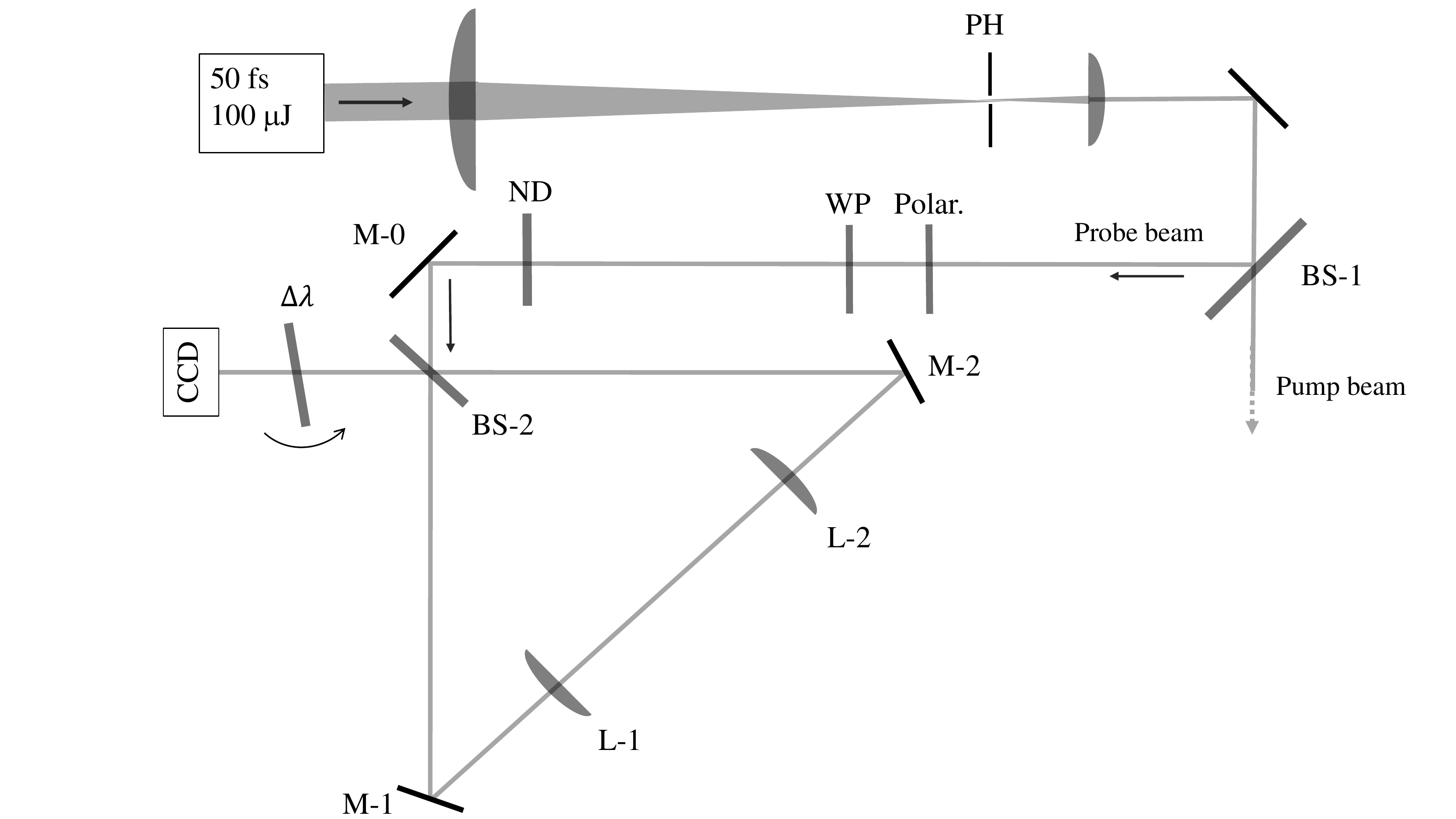}
  \caption{Schematic view of the DeLLight prototypes. Two different Sagnac interferometers have been developed and tested: on the left, a rectangular  interferometer composed of three mirrors, without any focus inside the interferometer; on the right, a triangular configuration composed of two mirrors and two optical lenses so as to focus the laser pulses at the midpoint (see text for details). }
  \label{fig:setup-prototype}
\end{figure}

\begin{figure}
\centering
\includegraphics[width=0.45\columnwidth]{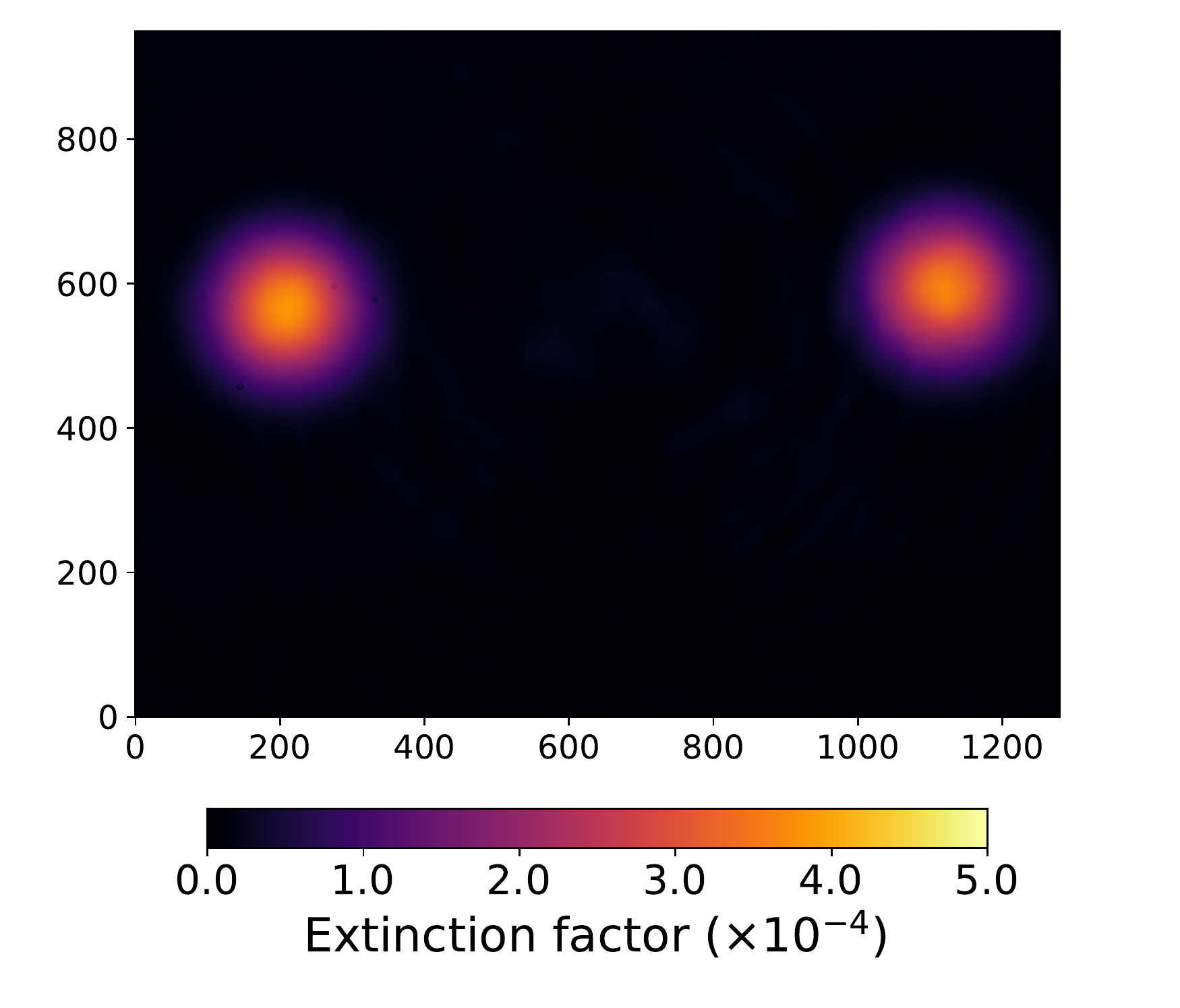} \, \includegraphics[width=0.45\columnwidth]{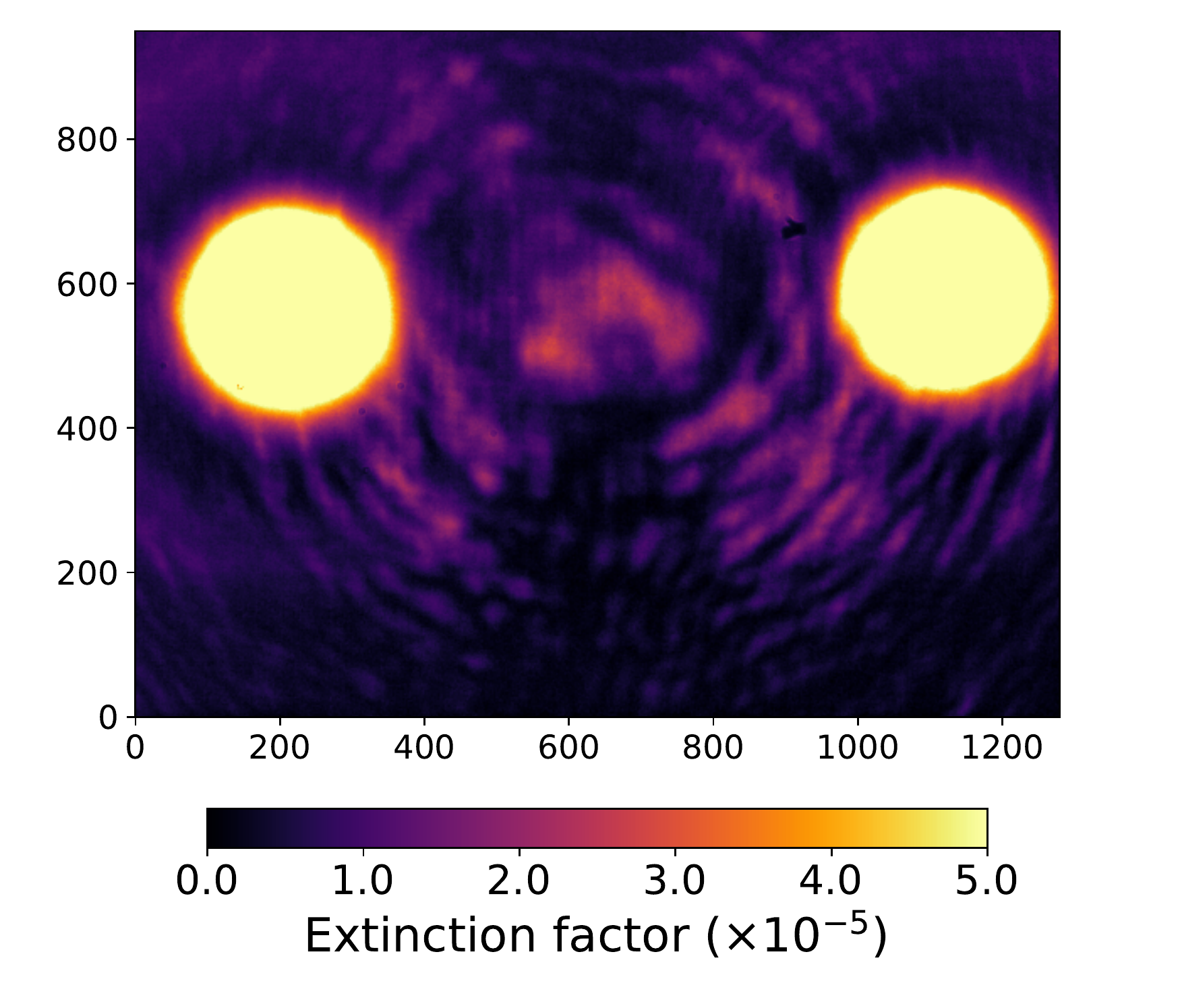}
  \caption{Intensity profile recorded by the CCD camera in the dark output of the interferometer at the maximum of extinction. The interference signal is located in the central part, delimited by the dotted white circle. The two opposite lateral spots correspond to the back reflections on the rear side of the beamsplitter BS-2 (See Figure~\ref{fig:setup-prototype}). On the right, same image shown with a higher sensitivity scale of the display in order to observe the residual phase noise of the interference signal with an extinction factor at worst equal to $2 \times 10^{-5}$. }
  \label{fig:extinction-max}
\end{figure}

\begin{figure}
\centering
  \includegraphics[scale=0.45]{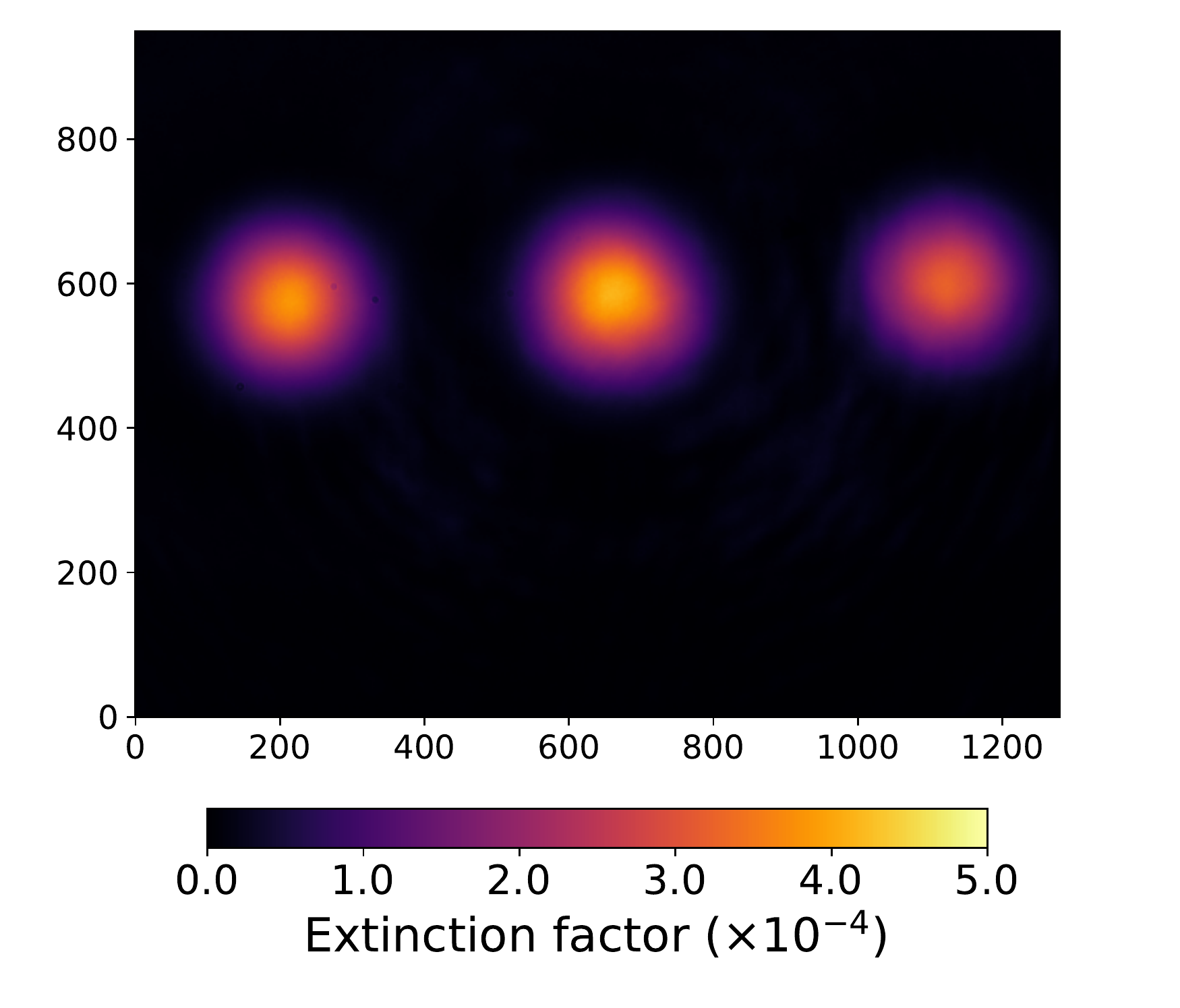}
  \caption{Intensity profile recorded by the CCD camera in the dark output of the interferometer, after having rotated the polarization of the probe beam at the entrance of the interferometer in order to have a signal intensity of the same order as the back-reflection intensities.} 
  \label{fig:rotated-polar}
\end{figure}

\begin{figure}
\centering
  \includegraphics[scale=0.65]{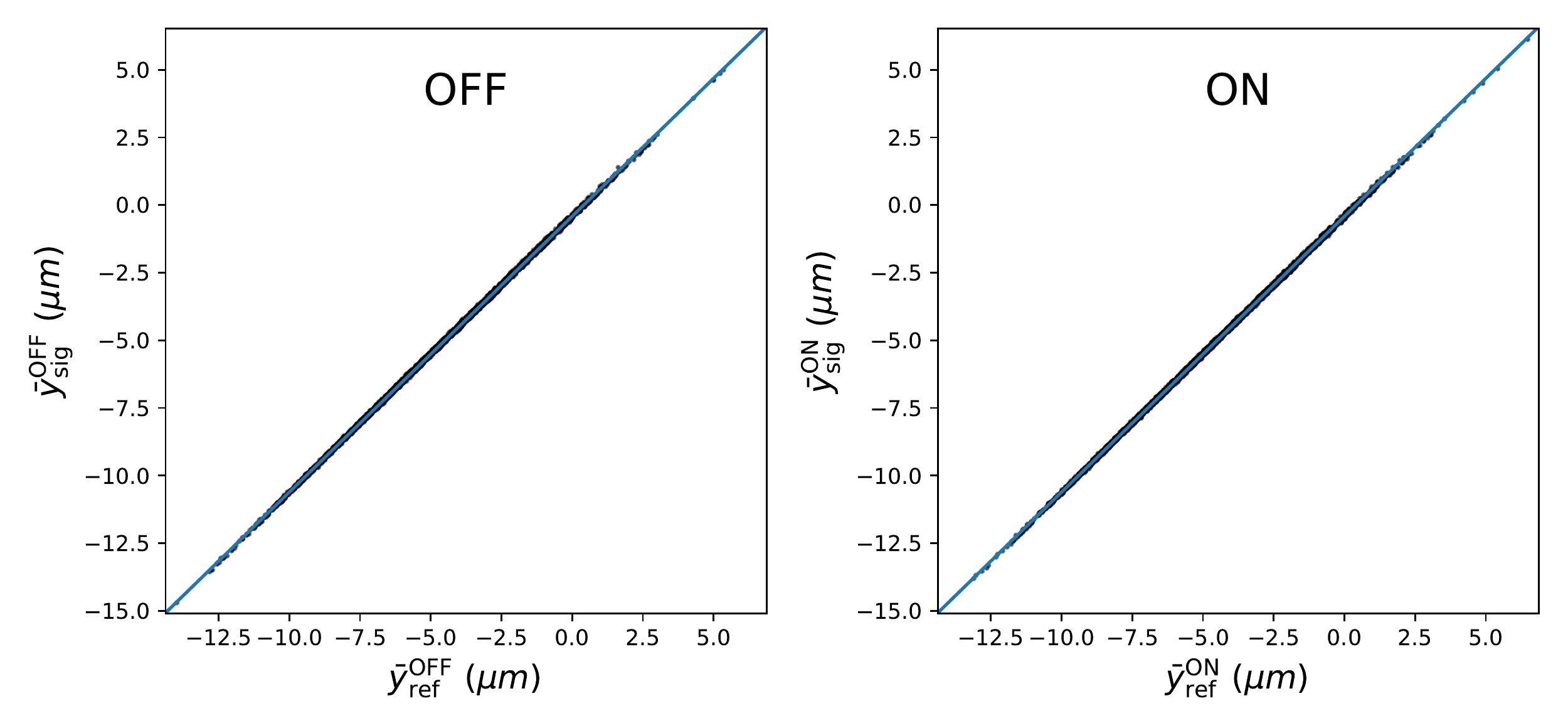}
  \caption{Correlation of the barycenters in intensity of the interference signal $\bar{y}_{\mathrm{sig}}(i)$ and the back-reflection $\bar{y}_{\mathrm{ref}}(i)$, calculated along the vertical axis $y$.  
The data is shown as black points, for both OFF (left panel) and ON (right panel) measurements, while the result of the linear fit obtained using the OFF data only is shown in both panels as a blue line.  The absence of any noticeable shift is consistent with the expected zero signal.}  
  \label{fig:beam-pointing-corrections}
\end{figure}

\begin{figure}
\centering
\includegraphics[width=0.49\columnwidth]{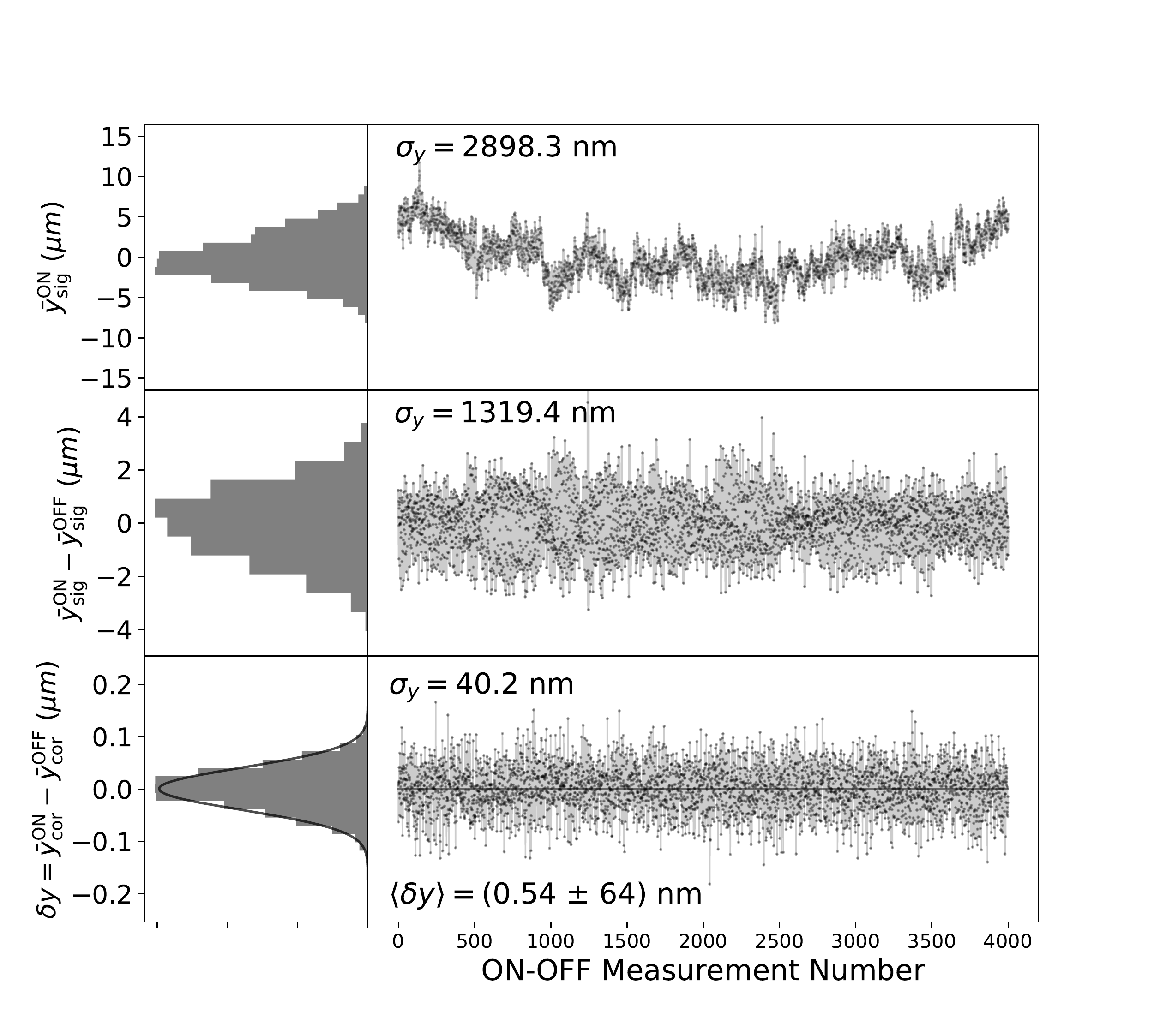} \,
\includegraphics[width=0.49\columnwidth]{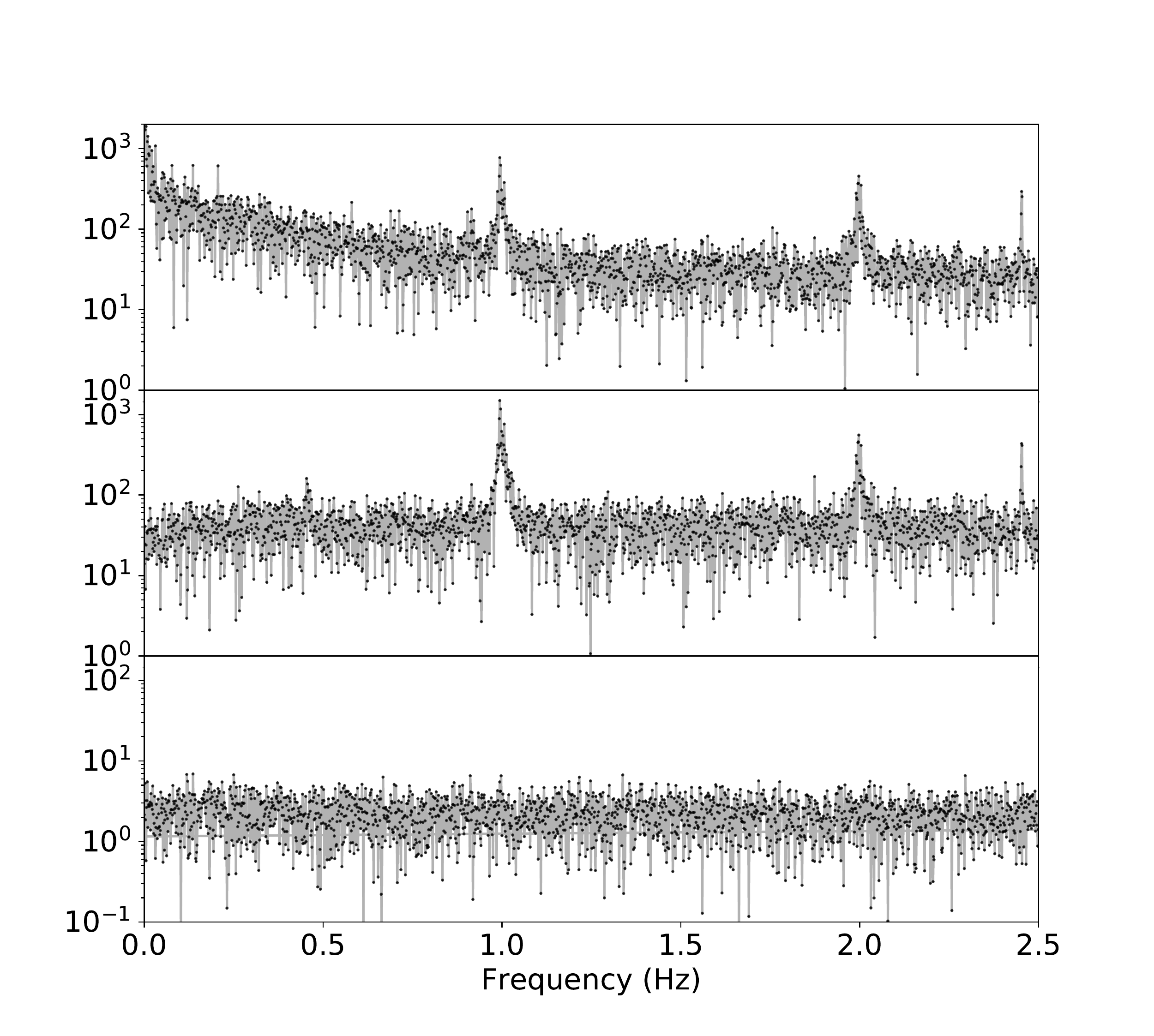}
\caption{Measurement of the spatial resolution obtained with 8000 successive laser shots at 10~Hz. (Left) Distribution of the barycenters in intensity of the interference signal as a function of the ``ON-OFF'' measurement $i$. Upper plot: raw barycenter position for the OFF data, $\bar{y}_{\mathrm{sig}}^{\mathrm{OFF}}(i)$, without any pointing correction; Middle plot: ``ON-OFF'' subtraction of the raw barycenter positions $\bar{y}_{\mathrm{sig}}^{\mathrm{ON}}(i) - \bar{y}_{\mathrm{sig}}^{\mathrm{OFF}}(i)$; Lower plot: corrected signal $\delta y (i)$, after beam pointing correction. Analysis is done with a region of interest of size $w_{\rm{RoI}} = w/2$, where $w$ is the width (fwhm) of the intensity profile of the beam. The achieved spatial resolution is $\sigma_y(w_{\rm{RoI}} = w/2) = 40.2 \pm 0.4$~nm and the average signal is $\langle{\delta y}\rangle = 540 \pm 636$~pm, which is compatible with the expected zero value. (Right) Corresponding frequency spectra. }
\label{fig:spatial-resolution}
\end{figure}

\begin{figure}
\centering
  \includegraphics[width=0.4\columnwidth]{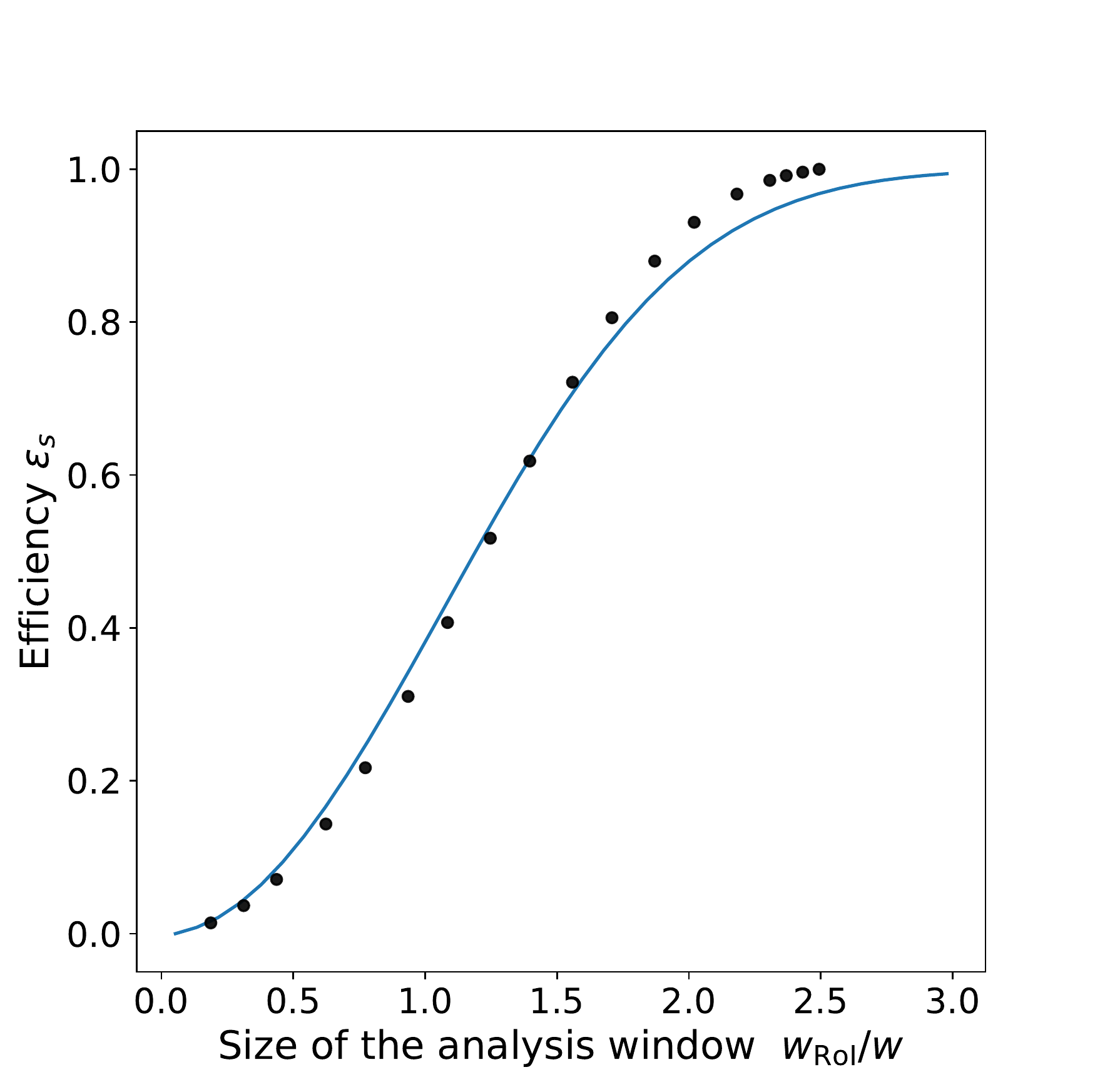} \,
  \includegraphics[width=0.49\columnwidth]{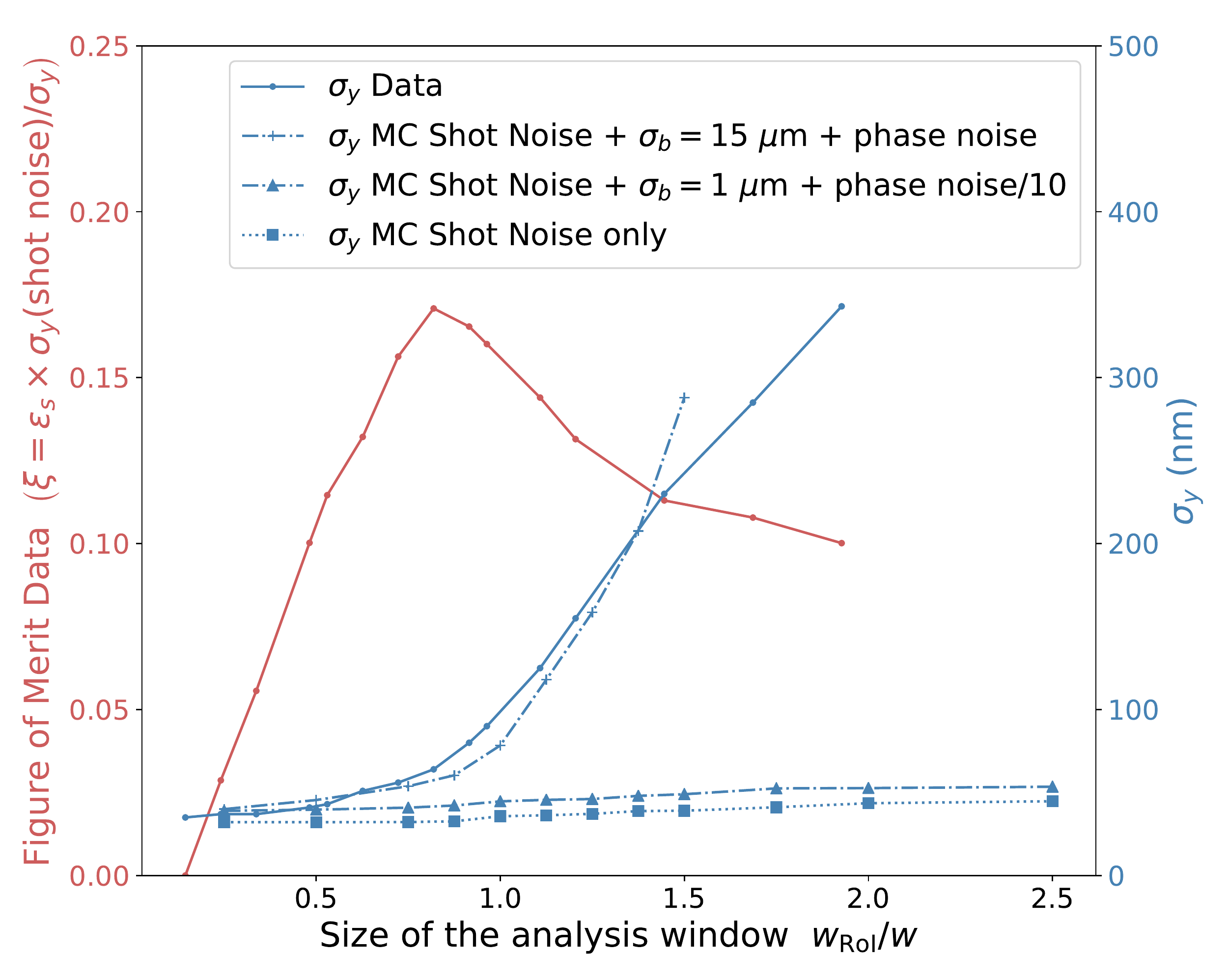}
  \caption{(Left) Efficiency $\epsilon_s$ of measuring the signal (i.e. the displacement of the barycenter) as a function of $w_{\rm RoI}/w$ where $w_{\rm RoI}$ is the size of the analysis window and $w$ the width (fwhm) of the beam, as measured in the data (dots) and calculated for a gaussian transverse profile of the beam. 
(Right) Spatial resolution $\sigma_y$ as a function of  $w_{\rm RoI}/w$. Solid blue line corresponds to the data and is compared to the Monte Carlo simulations with beam pointing fluctuations ($\sigma_b = 15 \mu$m) and a phase noise as measured in the data (crosses and dot dash line). Result of the simulations are also presented, assuming beam pointing fluctuations $\sigma_b = 1 \mu$m and a phase noise 10 times smaller (triangles and dot dash line), and assuming only shot noise (squares and dot line). The figure of merit defined as $ \xi(w_{\rm RoI}) = \epsilon_s(w_{\rm RoI}) \times \sigma_y(\mathrm{shot \ noise}) / \sigma_y (w_{\rm RoI}) $ is shown in red line. }
  \label{fig:fact-merite-sigy-roi}
\end{figure}

\begin{figure}
\includegraphics[width=0.3\columnwidth]{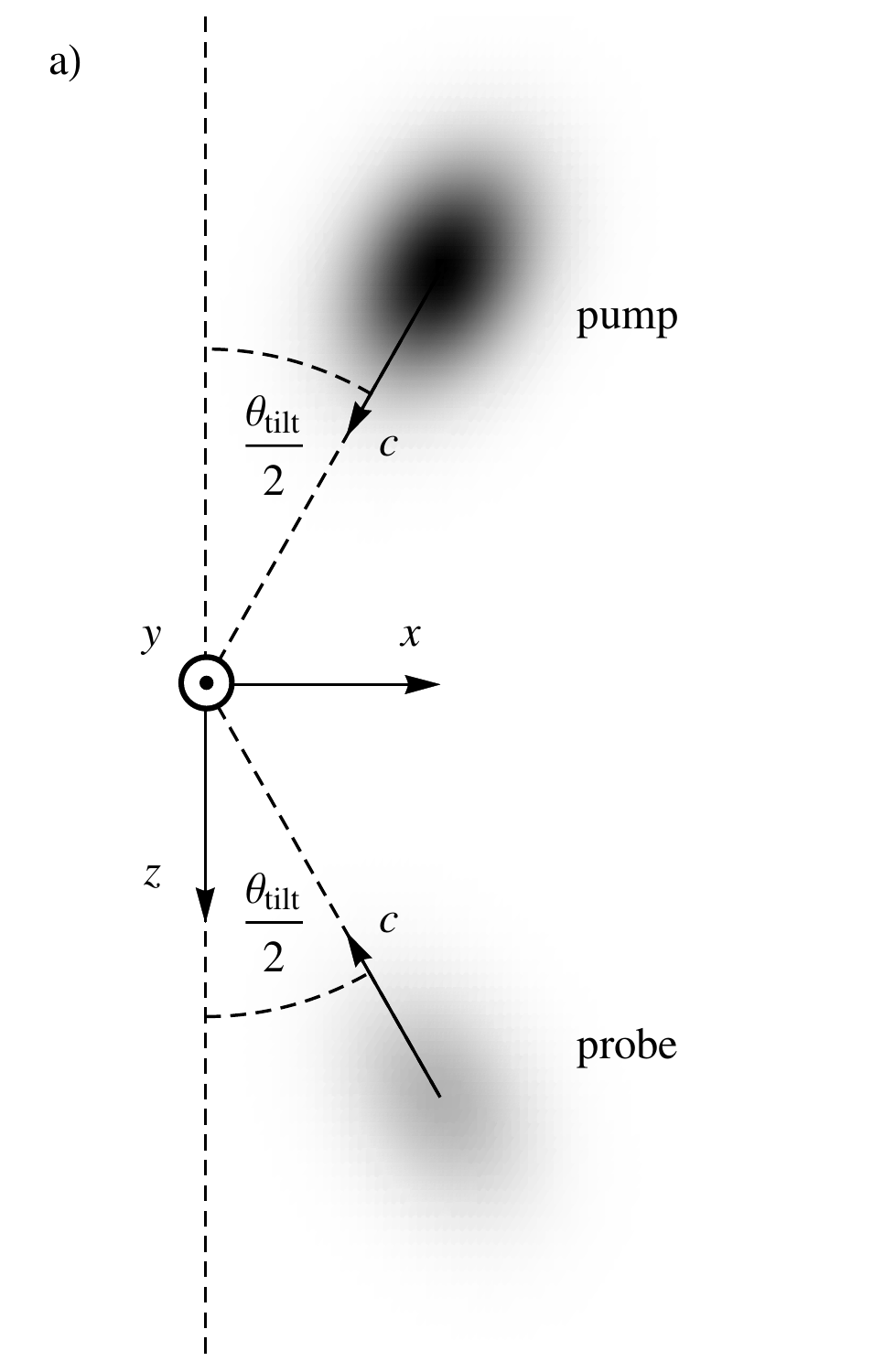} \, \includegraphics[width=0.3\columnwidth]{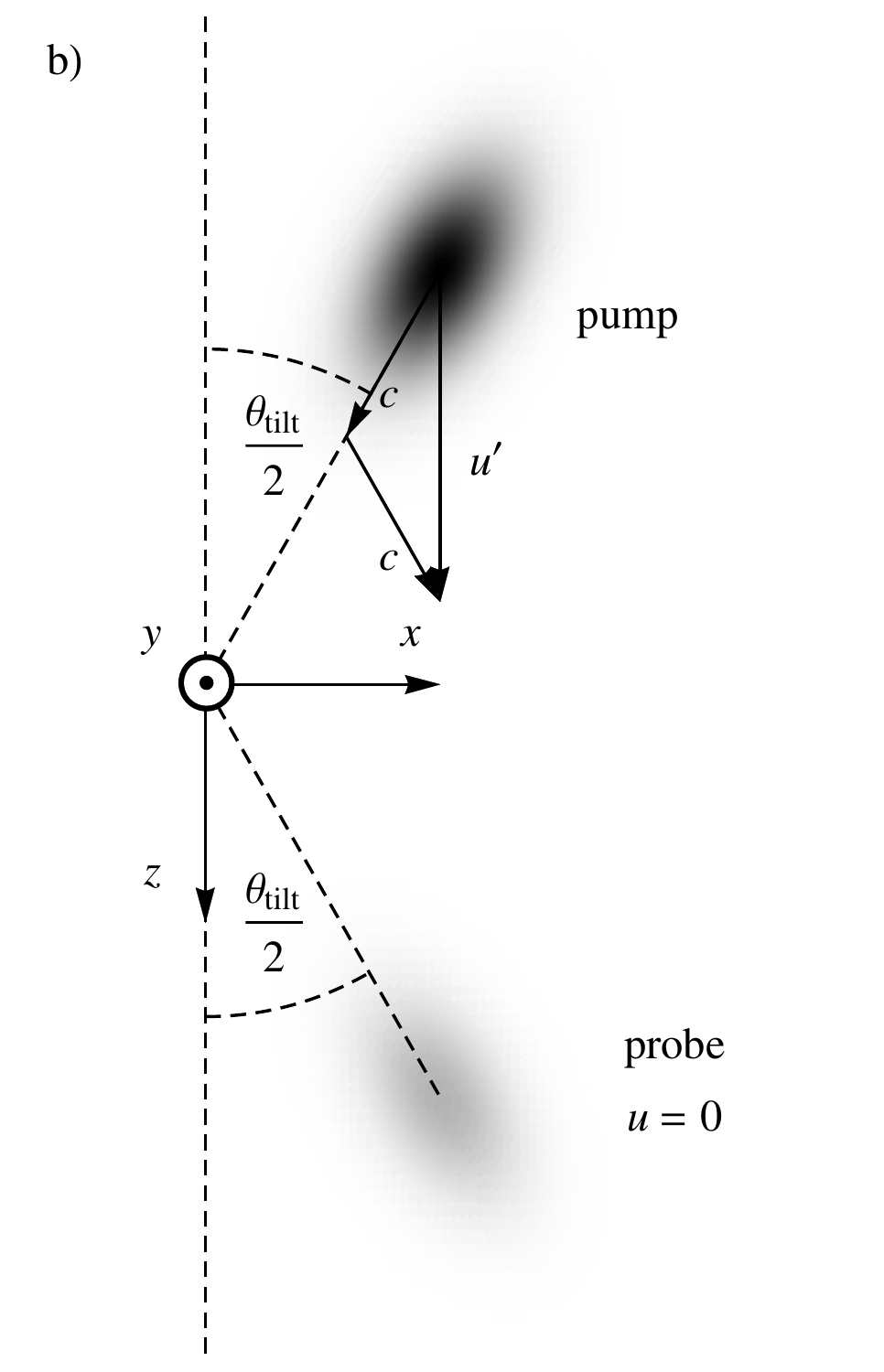}
\caption{Adopted coordinate system.   $a)$ In an inertial coordinate system which is fixed with respect to the lab, the wave vectors of pump and probe together span the $xz$-plane, and each makes an angle of $\theta_{\rm tilt}/2$ with the $z$-axis.  The total tilt angle between the two wave vectors is of course $\theta_{\rm tilt}$.  $b)$ Calculating the barycenter shift is most easily done in a coordinate system which is attached to the probe, related to that in $a)$ by subtracting the velocity of the probe.  In this coordinate system, the probe is of course stationary, while the pump moves towards it in the $z$-direction at a velocity $u^{\prime} = 2\,c\,{\rm cos}\left(\theta_{\rm tilt}/2\right)$.
\label{fig:pulses_angle}}
\end{figure}

\end{widetext}

\end{document}